\documentstyle[12pt,leqno,amsmath,amssymb,ifthen, url, graphicx,multirow,color,ulem,wrapfig, floatrow, subcaption,multicol]{article}
\newcommand{\mbf}[1]{\mbox{\boldmath $#1$}}
\newcommand{\ba}{{\mbf \beta}}

{\catcode `\@=11 \global\let\AddToReset=\@addtoreset}
\AddToReset{equation}{section}

\AddToReset{Theorem}{section}

\newtheorem{Def}{Definition}[section]
\newtheorem{lemma}{Lemma}[section]
\newtheorem{rem}{Remark}[section]
\newtheorem{theorem}{Theorem}[section]

\newcommand{\cD}{{\cal D}}

\newcommand{\cG}{{\cal G}}

\newcommand{\cL}{{\cal L}}

\newcommand{\cN}{{\cal N}}

\newcommand{\cS}{{\cal S}}
\newcommand{\cT}{{\cal T}}

\def\ba{\begin{array}}
\def\bc{\begin{center}}
\def\bd{\begin{description}}
\def\be{\begin{enumerate}}
\def\ea{\end{array}}
\def\ec{\end{center}}
\def\ed{\end{description}}
\def\edt{\end{document}}
\def\ee{\end{enumerate}}
\def\ben{\begin{equation}}
\def\benn{\begin{equation*}}
\def\een{\end{equation}}
\def\eenn{\end{equation*}}
\def\benr{\begin{eqnarray}}
\def\eenr{\end{eqnarray}}
\def\benrr{\begin{eqnarray*}}
\def\eenrr{\end{eqnarray*}}

\def\del{\delta}
\def\edt{\end{document}}

\def\iny{\infty}

\def\lel{\label}

\def\mb{\mbox}
\def\noi{\noindent}

\def\r{\ref}

\def\ra{\rightarrow}

\def\bfs{\bf s}

\def\si{\sigma}

\def\vep{\varepsilon}

\def\wh{\widehat}

\def\wti{\widetilde}

\def\N{{\mathbb N}}
\def\R{{\mathbb R}}

\def\bfS{{\mbf S}}
\def\bfs{{\mbf s}}

\begin{document}

\newcounter{num}
\setcounter{num}{0}

\newcounter{show}
\setcounter{show}{1}

\bc
{\bf \large A Fast Algorithm for Implementation of Koul's Minimum \\
Distance Estimators and Their Application to Image Segmentation\\
}
\ec

\bc
{\small{Jiwoong Kim\\
University of South Florida}\\
}
\ec

\begin{abstract}
\noi
Minimum distance estimation methodology based on an empirical distribution function has been popular due to its desirable properties including robustness. Even though the statistical literature is awash with the research on the minimum distance estimation, the most of it is confined to the theoretical findings: only few statisticians conducted research on the application of the method to real world problems. Through this paper, we extend the domain of application of this methodology to various applied fields by providing a solution to a rather challenging and complicated computational problem. The problem this paper tackles is an image segmentation which has been used in various fields. We propose a novel method based on the classical minimum distance estimation theory to solve the image segmentation problem. The performance of the proposed method is then further elevated by integrating it with the ``segmenting-together" strategy. We demonstrate that the proposed method combined with the segmenting-together strategy successfully completes the segmentation problem when it is applied to the complex real images such as magnetic resonance images.
\end{abstract}

\section{Introduction}
In a series of papers, Wolfowitz \cite{Wolfowitz1953}--
\cite{Wolfowitz1957} proposed a minimum distance (MD) estimation method for estimating the underlying parameters in some parametric models. The distances used in this  method are based on certain empirical processes. He showed that this method enables one to obtain consistent estimators under much weaker conditions than those required for the consistency of  maximum likelihood estimators. Much later, Donoho and Liu \cite{DONOHOD1988a}, \cite{DONOHOD1988b} argued that in the one- and two-sample location models the MD estimators based on integrated square distances involving residual empirical distribution functions have some desirable finite sample properties and tend to be automatically robust against some contaminated models. Koul \cite{Koul1985b},  \cite{Koul1985a}, \cite{Koul1986}, \cite{Koul2002} showed that in regression and autoregressive time series models, the  analogues of these estimators of the underlying parameters are given in terms of certain weighted residual empirical processes.
 These estimators include least absolute deviation (LAD), analogues of Hodges-Lehmann (H-L) estimators, and several other estimators that are robust against outliers in errors and asymptotically efficient at some error distributions. Kim \cite{KimJ2020} showed that the MD estimation can also be applied to a linear regression model with weakly dependent errors and empirically demonstrated the finite sample efficiency of some of these estimators. Despite these desirable properties, the application of the MD estimation methodology to real-world problems, however, stands in a nascent stage due to primarily the computational difficulty.  Still, the merits of the methodology demonstrated by many statisticians in the past decades are strong prima facie evidence that motivates the method to be applied to other problems. In this paper, we apply this methodology  to image segmentation problems with the hope that it will yield better inference in these problems.

One caveat against using the MD estimation method for solving the proposed image segmentation problem involves its computational complexity. One critical issue that causes this complexity for these estimators is illustrated in the next section. An important point to bear in mind is that naive application of existing numerical optimization methods to minimize an integrated square difference of weighted residual empirical processes will lead to a slow and even wrong computation. Rather than resorting to them, we, therefore, consummately investigate the structure of the distance function and exploit some of its useful characteristics to expedite the computation. To that end, we propose a novel algorithm to enable the application of this MD method to the image segmentation problem for the first time. The code used in this paper is available in GitHub repository: \url{https://github.com/jwboys26/Segmenting_Together}.

\section{Minimum distance estimators}
In this section, we discuss; (1) a regression model used in the image segmentation analysis; (2) MD estimators of the underlying parameters in the model; and (3) a fast computational algorithm to obtain those estimators. To be more precise we begin with some definitions and a model for the image segmentation.
\begin{Def}
In digital imaging, a \textbf{pixel} is a physical point in a raster image. A collection of $MN$ pixels is denoted by
\benn
S = \{(i,j): 1\leq i \leq M, 1\leq j\leq N\},
\eenn
where $M$ and $N$ are known positive integers.
\end{Def}
\noi
An image, $Img$, is a map from $S$ to the real line $\R$ that is expressed as an $M\times N$ matrix
\benn
Img := \left[
  \begin{array}{cccc}
    g(1,1) & g(1,2) & \cdots & g(1,N) \\
    g(2,1) & \cdots & \cdots & g(2,N) \\
    \vdots & \vdots & \vdots & \vdots \\
    g(M,1) & \cdots & \cdots & g(M,N) \\
  \end{array}
\right],
\eenn
where $g:\N^{2}\rightarrow \R^{D}$ and $g(i,j)$ denotes the value of the image at the $(i,j)$th pixel. Clearly, $Img=g(S)$. Here, $D$ denotes the number of channels. If $D=1$, the resultant output is a grayscale image while a color image that we commonly see corresponds to the case of $D=3$. This study considers grayscale images, i.e., $D=1$. Pixel values of the grayscale images represent the contrast that ranges from the darkest (black) to the brightest (white).

Let $K$ be a positive integer and $S_{k},\,1\le k\le K$ denote a partition of $S$ so that  $S_{i}\cap S_{j}=\emptyset$ for $i\neq j$ and
\benn
\bigcup_{k=1}^{K}S_{k} = S.
\eenn
Accordingly, let $n_{k}$ and $n$ denote cardinalities of $S_{k}$ and $S$, respectively so that $n=\sum_{k=1}^{K}n_{k}=MN$. Subsequently, we write
\benn
S_{k} = \{(x_{ik},y_{ik})\in N^{2}, \, i=1,2,...,n_{k}: 1\le x_{ik}\le M, 1\le y_{ik}\le N\},
\eenn
so that the image $Img$ can be segmented into $K$ sub-images: $g(S_{1})$, $g(S_{2})$, ..., $g(S_{K})$. For $1\le i \le n_{k}$ and $k=1,...,K$, let $S_{1}^{\dagger}, S_{2}^{\dagger},...,S_{K}^{\dagger}$ be a set of a mutually exclusive partition of $S$ by $K$ different colors and
\benn
g(x_{ik},y_{ik}) = p_{k}\quad \textrm{if }(x_{ik},y_{ik})\in S_{k}^{\dagger},
\eenn
where $0\le p_{k}\le 1$; black and white pixels take a value of 0 and 1, respectively, while a gray pixel takes an intermediate value.\\
\noi
\textbf{Example 1.} Consider $K=2$ and refer to Figure \r{fig:noise}. In the left panel of the figure, let $S_{1}^{\dagger}$ and $S_{2}^{\dagger}$ denote collections of white and black pixels, respectively. Note that a pair of $S_{1}^{\dagger}$ and $S_{2}^{\dagger}$ is a mutually exclusive, exhaustive partition of $S$ where $S$ is the collection of all pixels in the entire rectangle. Hence, for any pixel $\bfs\in S$,
\benn
g(\bfs)= \left\{
              \begin{array}{ll}
                1, & \hbox{if $\bfs\in S_{1}^{\dagger}$;} \\
                0, & \hbox{if $\bfs\in S_{2}^{\dagger}$.}
              \end{array}
            \right.
\eenn
In this example, $p_{1}$ and $p_{2}$ corresponding to
$S_{1}^{\dagger}$ and $S_{2}^{\dagger}$ will be 1 and 0, respectively.

For $1\le i \le n_{k}$ and $k=1,...,K$, define $\tilde{g}:\mathbb{N}^{2}\rightarrow \R$ by the following relation
\ben\lel{eq:Model}
\tilde{g}(x_{ik},y_{ik}):= g(x_{ik},y_{ik})+\vep_{i},
\een
where $\vep_{i}$ are some random variables. In the literature, $\vep_{i}$ is called ``noise" since it blurs the true pixel value $g(x_{ik},y_{ik})$. 
In the presence of noise, we will observe $\tilde{g}(\bfs)$ for $\bfs\in S$, instead of $g(\bfs)$: the right panel of Figure \r{fig:noise}  illustrates the presence of noise.

When no noise is observed as shown in the left panel of Figure \r{fig:noise}, separating white pixels from black pixels (or vice versa), i.e., identifying $S_{1}^{\dagger}$ and $S_{2}^{\dagger}$ does not foment any problems. In the right panel of Figure \r{fig:noise}, the same segmentation task -- separating the originally-white pixels from other pixels -- is encumbered by the presence of noise. Thus, the noise renders the segmentation task more challenging and error-prone: Figures \r{Fig:Wo_noise}-\r{Fig:noise_3} illustrate the results of the segmentation deteriorate even more as stronger noise is introduced to original images. For general $K$, the same conclusion holds. Therefore, the successful image segmentation hinges upon how to closely identify $S_{1}^{\dagger}, S_{2}^{\dagger},...,S_{K}^{\dagger}$ when they are blurred by noise, i.e., $\tilde{g}(S)$ is observed instead of $g(S)$ due to the presence of the noise. In the sequel, let $\bfs_{ik}:=(x_{ik},y_{ik})$ denote the $i$th pixel of $S_k$.
\ifnum \value{show}=1{
\begin{figure}[h]
    \centering
    \begin{subfigure}[b]{0.45\textwidth}
        \includegraphics[height=40mm, width=50mm]{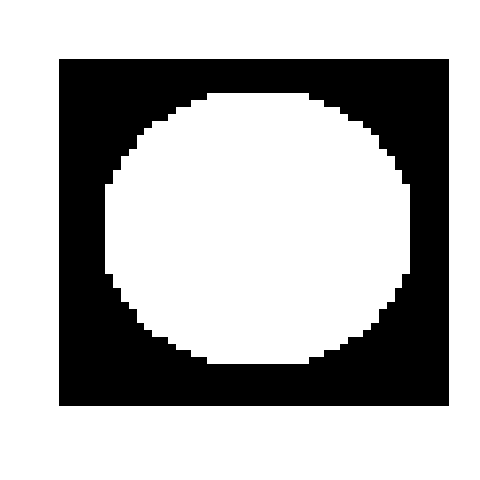}
    \end{subfigure}
    \begin{subfigure}[b]{0.45\textwidth}
        \includegraphics[height=40mm, width=50mm]{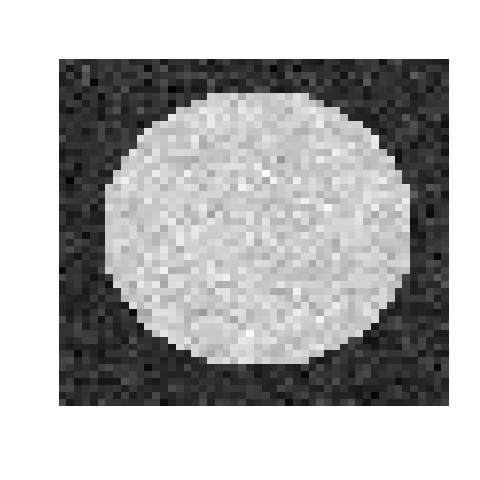}
    \end{subfigure}
    \caption{Images with and without a noise.}\lel{fig:noise}
\end{figure}
}\fi

Next, we shall define the distance function. Let
\benn
\boldsymbol{\cS}^{K} = \{(S_{1},S_{2},...,S_{K}): \cup_{i=1}^{K} S_{i} = S, \,S_{i}\cap S_{j}=\emptyset,
 \textrm{ for all }\,i\neq j \}.
\eenn
\noi
Note that any $\bfS\in \boldsymbol{\cS}^{K}$ will be then a $K$-tuple -- whose components are collections of pixels --
and partition $S$ into $K$ exhaustive, exclusive sets of pixels. Accordingly, define
\benn
\cL(\bfS) = \int \sum_{k=1}^{K} \left[ \frac{1}{n}\sum_{s_{i}\in S_{k}}  \Big\{ \textrm{I}\big(\tilde{g}(\bfs_{i}) - p_{k}\leq x \big) - \textrm{I}\big(-\tilde{g}(\bfs_{i}) + p_{k}< x \big)   \Big\} \right]^{2}dH(x),
\eenn
for any $\mbf{S}=(S_{1},...,S_{K})\in \boldsymbol{\cS}^{K}$, where $\textrm{I}(\cdot)$ is an indicator function, and $H$ is a nondecreasing right continuous function on $\R$ to $\R$ having left limits. The above class of distances, one for each $H$, is deduced from Koul \cite{Koul2002} which deals with parametric linear and nonlinear regression and autoregressive models. The image segmentation problem in terms of the above distance is equivalent to solving the minimization problem
\benn
\wh{\mbf{S}} = \mb{argmin}_{\bfs \in \boldsymbol{\cS}^{K}}\cL(\mbf{S}),
\eenn
where $\wh{\mbf{S}}=(\wh{S}_{1},...,\wh{S}_{K})$.

Before proceeding further, we need the following assumptions on $\vep_{i}$: $\vep_{i}$'s are independently and identically distributed according to $F$ with a finite second moment where $F$ is continuous and  symmetric around zero, i.e., $F(-x)=1-F(x)$ for all $x\in \R$. We do not, however, assume knowledge of this distribution.

\begin{rem}
In the context of the parametric linear regression model, the asymptotic normality of a large class of the analogues of the above minimum distance estimators is proved in Koul \cite{Koul2002}. For a given $F$ satisfying the given assumptions, it is possible to find an $H$ for which the corresponding MD estimator is asymptotically efficient. For example, if $F$ is the Laplace distribution then $H(x)\equiv \del_0(x)$ -- the distribution function degenerate at zero -- yields an asymptotically efficient MD estimator while $H(x)\equiv x$ gives an asymptotically efficient one for the logistic $F$. At the same time, both of these estimators are known to be robust against gross errors.
\end{rem}

Consider $K=2$ and $H(x)\equiv x$. Note that
\begin{align}
\cL(S_{1}, S_{2}) &= \frac{1}{n^{2}}\int \bigg[ \sum_{s_{i}\in S_{1}}\Big\{ \textrm{I}\big(\tilde{g}(\bfs_{i}) - p_{1}\leq x \big) - \textrm{I}\big(-\tilde{g}(\bfs_{i}) + p_{1}< x \big)   \Big\} \bigg]^{2}dx \lel{eq:obj} \\
                 &   \qquad + \frac{1}{n^{2}}\int\bigg[\sum_{s_{j}\in S_{2}}\Big\{ \textrm{I}\big(\tilde{g}(\bfs_{j}) - p_{2}\leq x \big) - \textrm{I}\big(-\tilde{g}(\bfs_{j}) + p_{2}< x \big)   \Big\}   \bigg]^{2}dx,\nonumber
\end{align}
where $S_{1}\cup S_{2}=S$ and $S_{1}\cap S_{2}=\emptyset$. Subsequently, define the MD estimator $(\wh{S}_{1}, \wh{S}_{2})$ as
\ben\lel{eq:opt}
\cL(\wh{S}_{1}, \wh{S}_{2}) = \inf_{(S_{1},S_{2})\in\boldsymbol{\cS}^{2}} \cL(S_{1},S_{2}).
\een
\begin{lemma}\lel{lem1}
$\cL$ is bounded in probability, that is, for all $\epsilon>0$ there is a $0<B_\epsilon<\iny$ and $N_{1\epsilon}$ such that
\benn 
P\big[\cL(S_{1},S_{2})\le B_\epsilon\big]\ge 1-\epsilon, \qquad \forall (S_{1}, S_{2})\in \boldsymbol{\cS}^{2}, \,\, \forall\, n\ge N_{1\epsilon}.
\eenn
Moreover, solutions to the optimization problem in (\r{eq:opt})  exist almost surely.
\end{lemma}
\textbf{Proof}. Arguing as in \cite[p.154-159]{Koul2002},
(\r{eq:obj}) can be simplified to
\benn  
\cL(S_{1},S_{2})= \frac{1}{n^{2}}\Bigg[\sum_{\bfs_{i}\in S_{1}}\sum_{\bfs_{j}\in S_{1}} f_{ij}^{1} + \sum_{\bfs_{i}\in S_{2}}\sum_{\bfs_{j}\in S_{2}}f_{ij}^{2}\Bigg],
\eenn
where
\benr\lel{eq:fij}
f_{ij}^{1}&=& |\tilde{g}(\bfs_{i})+ \tilde{g}(\bfs_{j})-2p_{1} | - |\tilde{g}(\bfs_{i})- \tilde{g}(\bfs_{j})|,\\
f_{ij}^{2}&=& |\tilde{g}(\bfs_{i})+ \tilde{g}(\bfs_{j})-2p_{2} | - |\tilde{g}(\bfs_{i})- \tilde{g}(\bfs_{j})|.\nonumber
\eenr
Note that for $k\in \{1,2\}$,
\ben\lel{eq:penalty}
|\tilde{g}(\bfs_{i})+ \tilde{g}(\bfs_{j})-2p_{k} |= \left\{
                                               \begin{array}{ll}
                                                 |\vep_{i}+\vep_{j}|, & \hbox{if $\bfs_{i},\bfs_{j}\in S_{k}^{\dagger}$;} \\
                                                 |\vep_{i}+\vep_{j}-2p_{k}+2p_{3-k}|, & \hbox{if $\bfs_{i},\bfs_{j}\in S_{3-k}^{\dagger}$;}\\
                                                 |\vep_{i}+\vep_{j}-p_{k}+p_{3-k}|, & \hbox{otherwise,} \\
                                               \end{array}
                                             \right.
\een
and
\benn
|\tilde{g}(\bfs_{i})- \tilde{g}(\bfs_{j})| = \left\{
                                               \begin{array}{ll}
                                                 |\vep_{i}-\vep_{j}|, & \hbox{if $\bfs_{i},\bfs_{j}\in S_{k}^{\dagger}$;} \\
                                                 |\vep_{i}-\vep_{j}|, & \hbox{if $\bfs_{i},\bfs_{j}\in S_{3-k}^{\dagger}$;}\\
                                                 |\vep_{i}-\vep_{j}+p_{k}-p_{3-k}|, & \hbox{if $\bfs_{i}\in S_{k}^{\dagger},\,\bfs_{j}\in S_{3-k}^{\dagger}$;} \\
                                                 |\vep_{i}-\vep_{j}-p_{k}+p_{3-k}|, & \hbox{if $\bfs_{i}\in S_{3-k}^{\dagger},\,\bfs_{j}\in S_{k}^{\dagger}$.} \\
                                               \end{array}
                                             \right.
\eenn
Therefore,
\benrr
E|\cL(S_{1},S_{2})| &\leq & \frac{1}{n^{2}}\sum_{s_{i}\in S_{1}}\sum_{s_{j}\in S_{1}} E\big[|\vep_{i}+ \vep_{j} | + |\vep_{i}- \vep_{j}|+2(p_{1}+p_{2})\big]\\
 &&\quad + \frac{1}{n^{2}}\sum_{s_{i}\in S_{2}}\sum_{s_{j}\in S_{2}}E\big[|\vep_{i}+ \vep_{j} | + |\vep_{i}- \vep_{j}|+2(p_{1}+p_{2})\big],\\
&\leq& 4 E|\vep_{1}|+2(p_{1}+p_{2})<\iny.
\eenrr
Then, the finite mean of the distance function and the Markov inequality readily imply the first claim. Since $\boldsymbol{\cS}^{2}$ is a finite set whose cardinality is $2^{n_{1}+n_{2}}$, any bounded function on it has a minimum, thereby completing the proof of the lemma.
\begin{rem}\lel{rem:penalty}
Consider the case of no noise: $\vep_{i}=0$ for all $i$. As shown in (\r{eq:penalty}), wrong identification of $S_{k}^{\dagger}$  will lead to an increase of $2|p_{1}-p_{2}|$ or $|p_{1}-p_{2}|$ in the distance function, which plays a role of a penalty for the wrong identification. When there is no noise, the optimal solution $(\wh{S}_{1}, \wh{S}_{2})$ will, therefore, completely overlap with $(S_{1}^{\dagger},S_{2}^{\dagger})$. The presence of noise, however, alter the modality of the optimization problem: the penalty will be weaker than it should be or even wrong. For example, consider the third case of (\r{eq:penalty}) with $\vep_{i}+\vep_{j}=p_{k}-p_{3-k}$: the penalty ($p_{k}-p_{3-k}$) is completely offset by the noise, and the wrong identification will not be punished at all. Consequently, the noise will drive the distance function to yield a solution that are different from $(S_{1}^{\dagger},S_{2}^{\dagger})$. This case serves to illustrate how the presence of noise renders the segmentation task more challenging.
\end{rem}
\noi
In general, a solution to the optimization problem (\r{eq:opt}) does not have any closed-form expressions,
and hence, we seek a solution through numerical optimization. To that end, we start with a pair of collections of randomly-selected pixels $(S_{1}^{(0)}, S_{2}^{(0)})$ in the initial stage, find a better pair of collections
$(S_{1}^{(1)}, S_{2}^{(1)})$ that yields a smaller value of $\cL$ than the value at the previous stage, and keep iterating this procedure until the convergence. Then, how can we select a better pair of collections? To answer
this question, we introduce a concept of a ``netgain"  which is the quintessential part of our proposed method. For
some pixel $\bfs\in S_{i}$, let $S_{i}^{(-\{\bfs\})}$ and $S_{i}^{(+\{\bfs\})}$ denote $S_{i}\backslash \{\bfs\}$
and $S_{i}\cup \{\bfs\}$, respectively.
\begin{Def}\lel{def:netgain}
Consider a pixel $\bfs_{i}\in S_{1}$. Then a netgain of $\bfs_{i}$ is defined as
\benn
NG(\mbf{s}_{i}; S_{1}):=\cL(S_{1}^{(-\{\bfs_{i}\})}, S_{2}^{(+\{\bfs_{i}\})})-\cL(S_{1}, S_{2}),
\eenn
which is a difference between two distance functions before and after transferring $\bfs_{i}$ from $S_{1}$ to $S_{2}$. Similarly, a netgain of $\bfs_{j}\in S_{2}$ is defined as
\benn
NG(\mbf{s}_{j}; S_{2}):=\cL(S_{1}^{(+\{\bfs_{j}\})}, S_{2}^{(-\{\bfs_{j}\})})-\cL(S_{1}, S_{2}).
\eenn
\end{Def}
\noi
Consider $\bfs_{k}\in S_{i}$. Recall that $S_{i}^{(-\{\bfs_{k}\})}$ and $S_{j}^{(+\{\bfs_{k}\})}$ are $S_{i}\backslash\{\bfs_{k}\}$ and $S_{j}\cup\{\bfs_{k}\}$, respectively. For the sake of brevity, let $S_{i}^{(-k)}$ and $S_{j}^{(+k)}$ denote these two sets, respectively. Analogously, for $\bfs_{k},\,\bfs_{h}\in S_{i}$, let $S_{i}^{(-k,h)}$ and $S_{j}^{(+k,h)}$ denote $S_{i}\backslash\{\bfs_{k},\bfs_{h}\}$ and $S_{j}\cup\{\bfs_{k},\bfs_{h}\}$, respectively. Recall $f_{ij}^{1}$ and $f_{ij}^{2}$ from (\r{eq:fij}). Note that for $\bfs_{k}\in S_{1}$
\benn
\sum_{s_{i}\in S_{1}}\sum_{s_{j}\in S_{1}} f_{ij}^{1} =  \sum_{s_{i}\in S_{1}^{(-k)}}\sum_{s_{j}\in S_{1}^{(-k)}} f_{ij}^{1}+ 2\sum_{s_{i}\in S_{1}^{(-k)}} f_{k i}^{1} +  f_{kk}^{1},
\eenn
and
\benn
\sum_{s_{i}\in S_{2}}\sum_{s_{j}\in S_{2}} f_{ij}^{2}=\sum_{s_{i}\in S_{2}^{(+k)}}\sum_{s_{j}\in S_{2}^{(+k)}} f_{ij}^{2}- 2\sum_{s_{i}\in S_{2}} f_{k i}^{2} -  f_{kk}^{2}.
\eenn
Therefore,
\benr
NG(\bfs_{k};S_{1})&=& \cL(S_{1}^{(-k)}, S_{2}^{(+k)}) - \cL(S_{1}, S_{2}),\lel{eq:NG} \\
            &=& \frac{1}{n^{2}}\left[-2\sum_{s_{i}\in S_{1}}f_{k i}^{1}+2\sum_{s_{j}\in S_{2}}f_{k j}^{2} + f_{kk}^{1}+ f_{kk}^{2}\right].\nonumber
\eenr
\noi
$E|\vep_{i}|<\iny$ readily implies $NG(\bfs_{k};S_{1})$ is stochastically bounded, i.e., $NG(\bfs_{k};S_{1})=O_{p}(1)$. Adoption of the concept of the netgain makes it convenient to analyze the distance function in that any difference between distance functions -- one obtained from the original $(S_{1}, S_{2})$ and another obtained after the transfer of several pixels from $S_{1}$ to $S_{2}$ -- can be written as the sum of netgains of those pixels. For example, we have
\benr
\cL(S_{1}^{(-k,h)}, S_{2}^{(+k,h)}) - \cL(S_{1}, S_{2})
&=& NG(\bfs_{k};S_{1}) + NG(\bfs_{h};S_{1}^{(-k)}),\lel{eq:loss_diff}\\
&=&NG(\bfs_{h};S_{1}) + NG(\bfs_{k};S_{1}^{(-h)})\nonumber,
\eenr
for $\bfs_{k},\bfs_{h}\in S_{1}$. Similarly,
\benrr
\cL(S_{1}^{(+k,h)}, S_{2}^{(-k,h)}) - \cL(S_{1}, S_{2})
&=& NG(\bfs_{k};S_{2}) + NG(\bfs_{h};S_{2}^{(-k)}),\\
&=&NG(\bfs_{h};S_{2}) + NG(\bfs_{k};S_{2}^{(-h)}),
\eenrr
when $\bfs_{k},\bfs_{h}\in S_{2}$. For the general case, consider any $S_{1}^{sub}\subset S_{1}$ where $S_{1}^{sub}=S_{1}\backslash\{\mbf{u}_{1},\mbf{u}_{2},...,\mbf{u}_{L}\}$. The difference of the distance functions before and after transfer of these $L$ pixels from $S_{1}$ to $S_{2}$ can be expressed as
\benn
\cL(S_{1}^{sub}, S\backslash S_{1}^{sub})-\cL(S_{1},S_{2})
= NG(\mbf{u}_{1};S_{1})+\sum_{i=2}^{L}NG(\mbf{u}_{i};S_{1}\backslash\cup_{j=1}^{i-1}\{\mbf{u}_{j}\}).
\eenn
\noi
For further discussion, define two sets:
\benn
\cN\cG(S_{1}):= \{\bfs_{i}\in S_{1}: NG(\bfs_{i};S_{1})< 0\},\,\,
\cN\cG(S_{2}):= \{\bfs_{j}\in S_{2}: NG(\bfs_{j};S_{2})< 0\}.
\eenn
From the definition of the netgain, $\cN\cG(S_{1})$ is a set of pixels which originally belong to $S_{1}$, and whose relocation from $S_{1}$ to $S_{2}$ will decrease the distance function. Similarly, $\cN\cG(S_{2})$ is a set of pixels whose relocation in the opposite way will decrease the distance function.

At this juncture, one important question arises: which pixels of $S_{1}$ and $S_{2}$ should be relocated to each other in order to decrease the distance function? It is not unreasonable to consider pixels which belong to $\cN\cG(S_{1})$ and $\cN\cG(S_{2})$ as potential candidates. For the convenience of further analysis, let $\cN\cG(S_{1})=\{\bfs_{1}^{1}, \bfs_{2}^{1},...,\bfs_{m_{1}}^{1}\}$ and
$\cN\cG(S_{2})=\{\bfs_{1}^{2}, \bfs_{2}^{2},...,\bfs_{m_{2}}^{2}\}$.
Transferring all elements of $\cN\cG(S_{1})$ from $S_{1}$ to $S_{2}$ will result in two new sets: $S_{1}\backslash \cN\cG(S_{1})$ and $S_{2}\cup \cN\cG(S_{1})$. Then, the difference between the consequential and initial distance functions can be written as
\benn
\cL(S_{1}\backslash\cN\cG(S_{1}), S_{2}\cup\cN\cG(S_{1})) - \cL(S_{1}, S_{2})
=NG(\bfs_{1}^{1};S_{1})+\sum_{i=2}^{m_{1}}NG\left(\bfs_{i}^{1};S_{1}\backslash\cup_{j=1}^{i-1}\{\bfs_{j}^{1}\}\right).
\eenn
It is worth noting the following facts; (1) $\bfs_{1}^{1}\in \cN\cG(S_{1})$ implies that the first term in the right-hand side of the above equation is less than 0, which means the transfer of $\bfs_{1}^{1}$ will always decrease the distance function; (2) the summand of the right-hand side of the equation $NG(\bfs_{i}^{1};S_{1}\backslash\cup_{j=1}^{i-1}\{\bfs_{j}^{1}\})$ is not always negative even though $NG(\bfs_{i}^{1};S_{1})<0$; and (3) transfer of all $\bfs_{i}^{1}$ in $\cN\cG(S_{1})$, hence, does not guarantee the greatest decrease in the distance function. To accomplish the desired result, define $\cT(S_{1})=\{\mbf{t}_{1},\mbf{t}_{2},...,\mbf{t}_{m_{1}^{*}}\}\subset \cN\cG(S_{1})$ as follows:
\benr
&&\mbf{t}_{1}:=\mb{argmin}_{\bfs\in \cN\cG(S_{1})}NG(\bfs;S_{1}),\lel{eq: TS1}\\
&&\mbf{t}_{2}:=\mb{argmin}_{\bfs\in \cN\cG(S_{1})} NG(\bfs;S_{1}\backslash\{\mbf{t}_{1}\}) \textrm { subject to } NG(\bfs;S_{1}\backslash\{\mbf{t}_{1}\})<0,\nonumber\\
&&\quad\quad \vdots \nonumber\\
&&\mbf{t}_{m_{1}^{*}}:=\mb{argmin}_{\bfs\in \cN\cG(S_{1})}NG(\bfs;S_{1}\backslash\cup_{j=1}^{m_{1}^{*}-1}\{\mbf{t}_{j}\}) \nonumber\\
&&\qquad\qquad\qquad\qquad \textrm { subject to } NG(\bfs;S_{1}\backslash\cup_{j=1}^{m_{1}^{*}-1}\{\mbf{t}_{j}\})<0.\nonumber
\eenr
To be more precise, we construct $\cT(S_{1})$ as follows. Among pixels of $\cN\cG(S_{1})$, find one which has the smallest netgain $NG(\bfs;S_{1})$; this pixel will be the first element of $\cT(S_{1})$, denoted by $\mbf{t}_{1}$. Note that $NG(\mbf{t}_{1};S_{1})<0$ by the definition of $\cN\cG(S_{1})$. Next, among the remaining pixels after removing $\mbf{t}_{1}$ from $\cN\cG(S_{1})$, find $\mbf{t}_{2}$ which has the smallest $NG(\bfs;S_{1}/\{\mbf{t}_{1}\})$, and check the sign of the resulting netgain which will play a role of a stopping criterion. If $NG(\mbf{t}_{2};S_{1}/\{\mbf{t}_{1}\})\geq0$, the construction of $\cT(S_{1})$ will be halted; otherwise, we add it to $\cT(S_{1})$ and proceed to find $\mbf{t}_{3}$. Repeating this procedure until the stopping criterion is met will yield $\cT(S_{1})$. As a result, $m_{1}^{*}\leq m_{1}$. $\cT(S_{2})\subset \cN\cG(S_{2})$ can be obtained in the same manner. By construction, we always have $\cL(S_{1}/\cT(S_{1}),S_{2}\cup\cT(S_{1}))<\cL(S_{1},S_{2})$ and $\cL(S_{1}\cup\cT(S_{2}),S_{2}\backslash\cT(S_{2}))<\cL(S_{1},S_{2})$. At the moment of trading some elements between $S_{1}$ and $S_{2}$, we will, therefore, refer to $\cT(S_{1})$ and $\cT(S_{2})$ instead of $\cN\cG(S_{1})$ and $\cN\cG(S_{2})$. Before proceeding further, define a function $sgn:\R\ra\R$ as follows:
\benn
sgn(x):=\left\{
                   \begin{array}{ll}
                     1, & \hbox{if $x\geq 0$;} \\
                     -1, & \hbox{if $x<0$.}
                   \end{array}
                 \right.
\eenn
\begin{lemma}\lel{eq:NG_positive}
For all $\epsilon>0$, there is a $N_{2\epsilon}$ such that
\benn
P\Big[ sgn( NG( \bfs_{h}; S_{1}^{(-k)}))= sgn(NG(\bfs_{h};S_{1}))\Big]\geq 1-\epsilon, \,\,\forall \bfs_{k}, \bfs_{h}\in S_{1},\,\forall n\geq N_{2\epsilon}.
\eenn
\end{lemma}
\begin{rem}
Lemma \r{eq:NG_positive} states that  $NG(\bfs_{h};S_{1})$ and $NG(\bfs_{h};S_{1}^{(-k)})$ have the same sign in probability. When constructing $\cT(S_{1})$, we will encounter the following question:  is it sufficient to consider pixels belonging $\cN\cG(S_{1})$ only when we choose another element of $S_{1}$ after selecting $\bfs_{k}\in S_{1}$. Lemma \r{eq:NG_positive} shows that $\bfs_{h}\in S_{1}$ does not need to be considered a candidate for the element of $\cT(S_{1})$ if $\bfs_{h}\notin \cN\cG(S_{1})$. Therefore, checking elements of $\cN\cG(S_{1})$ only will be good enough to construct $\cT(S_{1})$: the analogous result holds true for $S_{2}$. 
\end{rem}
\noi
\textbf{Proof.} Note that $\bfs_{h}\notin \cN\cG(S_{1})$ implies $NG(\bfs_{h}; S_{1})\geq0$ while the opposite holds when $\bfs_{h}\in \cN\cG(S_{1})$. It is not difficult to see (\r{eq:NG}) together with (\r{eq:loss_diff}) yields
\benn
NG(\bfs_{h};S_{1}^{(-k)})
= NG(\bfs_{h}; S_{1})+2n^{-2}(f_{h k}^{1}+f_{h k}^{2}).
\eenn
From (\r{eq:fij}), it follows that the second term in the above equation is $o_{p}(1)$. Thus, the claim follows from $NG(\bfs_{h}; S_{1})=NG(\bfs_{h}; S_{1}^{(-k)})=O_{p}(1)$.
\begin{lemma}\lel{cor1}
For $\bfs_{k},\bfs_{h}, \bfs_{l}\in S_{1}$, $NG(\bfs_{k};S_{1})>NG(\bfs_{h};S_{1})$ implies for all $\epsilon>0$ there is a $N_{3\epsilon}$ such that
\benn
P\big[ NG(\bfs_{k};S_{1}^{(-l)})>NG(\bfs_{h};S_{1}^{(-l)})\big]\geq1-\epsilon,\,\,\forall n\geq N_{3\epsilon}.
\eenn
The reverse conclusion also holds true, i.e.,    $NG(\bfs_{k};S_{1}^{(-l)})>NG(\bfs_{h};S_{1}^{(-l)})$ implies for all $\epsilon>0$ there is a $N_{4\epsilon}$ such that
\benn
P\big[ NG(\bfs_{k};S_{1})>NG(\bfs_{h};S_{1})\big]\geq 1-\epsilon,\,\,\forall n\geq N_{4\epsilon}.
\eenn
\end{lemma}
\noi
\textbf{Proof}. Akin to the previous lemma, observe that from (\r{eq:NG}) and (\r{eq:loss_diff})
\benrr
NG(\bfs_{k};S_{1}^{(-l)})-NG(\bfs_{h};S_{1}^{(-l)})&=& NG(\bfs_{k};S_{1})-NG(\bfs_{h};S_{1})\\
&& + n^{-2}(f_{l k}^{1}+f_{l k}^{2}) - n^{-2}(f_{l h}^{1}+f_{l h}^{2}).
\eenrr
The fact that both $n^{-2}(f_{l k}^{1}+f_{l k}^{2})$ and $n^{-2}(f_{l h}^{1}+f_{l h}^{2})$ are $o_{p}(1)$ completes the proof of the claim.\\
\noi
Starting with an initial pair of sets $(S_{1}, S_{2})$, we compute
$\cT(S_{1})$ and transfer its elements to $S_{2}$; let $\wti{S}_{1}=S_{1}\backslash\cT(S_{1})$ and $\wti{S}_{2}=S_{2}\cup\cT(S_{1})$. Now we compute $\cT(\wti{S}_{2})$, i.e., we find $\bfs\in \wti{S}_{2}$ whose netgain is less than 0 and transfer of which to $\wti{S}_{1}$ will decrease the distance function. When checking pixels belonging to $\wti{S}_{2}$, we may need to check the pixels initially belonging to $S_{2}$ only, which seems quite reasonable. To put it another way, any elements of $\cT(S_{1})$ may not be relocated back to $S_{1}$ once they move to $S_{2}$. The justification for this argument is primarily based on the following lemma.
\begin{lemma}\lel{lem:TS2}
For all $\epsilon>0$, there is a $N_{5\epsilon}$ such that
\benn
P\big[NG(\bfs_{k};\wti{S}_{2})> 0\big]>1-\epsilon, \quad \forall \bfs_{k}\in \cT(S_{1}),\,\, \forall n\geq N_{5\epsilon}.
\eenn
\end{lemma}
\begin{rem}
Once some elements of $S_{1}$ are transferred to $S_{2}$, they will not be relocated back to $S_{1}$. Lemma \r{lem:TS2}, therefore, implies remarkable curtailment of the computational cost; otherwise, their netgains should be redundantly computed. This redundancy will incur cumbersome computation, especially in the initial stage of the computation when the cardinality of $\cT(S_{1})$ is usually large.
\end{rem}
\noi
\textbf{Proof}. Assume that $\cT(S_{1})=\{\bfs_{1}^{1},\bfs_{2}^{1},...,\bfs_{h}^{1}\}$ and $\bfs_{i}^{1}$ was chosen at the $i$th stage of the scheme that is stated in (\r{eq: TS1}). To begin with, we will prove the case of $h=3$ and show that netgains of those three pixels are all positive in probability. When checking the netgains, we try in reverse order, thereby checking a netgain of $\bfs_{3}^{1}$ first and that of $\bfs_{1}^{1}$ last. The reason for this is that the difficulty for the proof of the claim dramatically decreases in the reverse order. Since we transferred $\bfs_{3}^{1}$ last from $S_{1}$ to $S_{2}$, $NG(\bfs_{3}^{1};\wti{S}_{2})$ is simply equal to $-NG(\bfs_{3}^{1};S_{1}\backslash\{\bfs_{1}^{1},\bfs_{2}^{1}\})$ that is strictly positive; otherwise, $\bfs_{3}^{1}\notin \cT(S_{1})$ in that we would not transfer $\bfs_{3}^{1}$ from $S_{1}$ to $S_{2}$ after $\bfs_{1}^{1}$ and $\bfs_{2}^{1}$ were transferred. Therefore, the claim for $\bfs_{3}^{1}$ holds.

Next, consider the netgain of $\bfs_{2}^{1}$. Observe that
\benrr
&&NG(\bfs_{2}^{1};\wti{S}_{2})\\
&=&\cL(S_{1}\backslash\{\bfs_{1}^{1}, \bfs_{3}^{1}\},S_{2}\cup\{\bfs_{1}^{1},\bfs_{3}^{1}\})-\cL(\wti{S}_{1},\wti{S}_{2}),\\
&=&\Big[\cL(S_{1}\backslash\{\bfs_{1}^{1}, \bfs_{3}^{1}\},S_{2}\cup\{\bfs_{1}^{1}, \bfs_{3}^{1}\})-\cL(S_{1},S_{2})\Big]- \Big[\cL(\wti{S}_{1},\wti{S}_{2})-\cL(S_{1},S_{2})\Big],\\
&=& \Big[NG(\bfs_{1}^{1};S_{1})+NG(\bfs_{3}^{1};S_{1}\backslash\{\bfs_{1}^{1}\})\Big]- \Big[NG(\bfs_{1}^{1};S_{1})+NG(\bfs_{2}^{1};S_{1}\backslash\{\bfs_{1}^{1}\}), \\
&&\quad\quad + NG(\bfs_{3}^{1};S_{1}\backslash\{\bfs_{1}^{1},\bfs_{2}^{1}\}) \Big],\\
&=& \Big[NG(\bfs_{3}^{1};S_{1}\backslash\{\bfs_{1}^{1}\})-NG(\bfs_{2}^{1};S_{1}\backslash\{\bfs_{1}^{1}\})\Big]
-NG(\bfs_{3}^{1};S_{1}\backslash\{\bfs_{1}^{1},\bfs_{2}^{1}\}),\\
&>&0,
\eenrr
where the inequality follows from  $NG(\bfs_{3}^{1};S_{1}\backslash\{\bfs_{1}^{1},\bfs_{2}^{1}\})=-NG(\bfs_{3}^{1};\wti{S}_{2})<0$ -- which was already proven -- and the way that $\cT(S_{1})$ is constructed as stated in (\r{eq: TS1}), i.e., $NG(\bfs_{2}^{1};S_{1}\backslash\{\bfs_{1}^{1}\})\leq NG(\bfs_{3}^{1};S_{1}\backslash\{\bfs_{1}^{1}\})$: otherwise $\bfs_{2}^{1}$ couldn't be chosen as the second element of $\cT(S_{1})$.

Finally, consider the netgain of $\bfs_{1}^{1}$. Note that $\cL(\wti{S}_{1},\wti{S}_{2})-\cL(S_{1},S_{2})$ in the previous equation can be rewritten as
\benrr
\cL(\wti{S}_{1},\wti{S}_{2})-\cL(S_{1},S_{2}) &=& NG(\bfs_{2}^{1};S_{1})+NG(\bfs_{1}^{1};S_{1}\backslash\{\bfs_{2}^{1}\})
 + NG(\bfs_{3}^{1};S_{1}\backslash\{\bfs_{1}^{1},\bfs_{2}^{1}\}),
\eenrr
which, in turn, yields
\benrr
NG(\bfs_{1}^{1};\wti{S}_{2})
&=& \Big[NG(\bfs_{2}^{1};S_{1})+NG(\bfs_{3}^{1};S_{1}\backslash\{\bfs_{2}^{1}\})\Big]- \Big[NG(\bfs_{2}^{1};S_{1})+NG(\bfs_{1}^{1};S_{1}\backslash\{\bfs_{2}^{1}\})\\
&&\quad\quad + NG(\bfs_{3}^{1};S_{1}\backslash\{\bfs_{1}^{1},\bfs_{2}^{1}\}) \Big],\\
&=&\Big[NG(\bfs_{3}^{1};S_{1}\backslash\{\bfs_{2}^{1}\})-NG(\bfs_{1}^{1};S_{1}\backslash\{\bfs_{2}^{1}\})\Big]
 -NG(\bfs_{3}^{1};S_{1}\backslash\{\bfs_{1}^{1},\bfs_{2}^{1}\}).
\eenrr
Observing  $NG(\bfs_{3}^{1};S_{1}) > NG(\bfs_{1}^{1};S_{1})$ from (\r{eq: TS1}), Lemma \r{cor1} readily implies the first term is positive in probability. This fact together with $NG(\bfs_{3}^{1};S_{1}\backslash\{\bfs_{1}^{1},\bfs_{2}^{1}\})<0$, in turn, implies $NG(\bfs_{1}^{1};\wti{S}_{2})>0$ in probability, thereby completing the proof of the case of $h=3$.

The proof for the general case ($h>3$) is very similar to the case of $h=3$, albeit a bit complicating. As in the case of $h=3$, the netgain of the last pixel $\bfs_{h}^{1}$, $NG(\bfs_{h}^{1};\wti{S}_{2})$, is strictly positive: otherwise, $\bfs_{h}^{1}\notin \cT(S_{1})$. Consider $NG(\bfs_{j}^{1};\wti{S}_{2})$,  $1<j<h$. Let $\cT_{l}^{1}:=\{\bfs_{1}^{1},...,\bfs_{l}^{1}\}\subset \cT(S_{1})$: e.g., $\cT_{h}^{1}= \cT(S_{1})$. Also, define $\cT_{l, -j}^{1}:=\cT_{l}^{1}\backslash \{\bfs_{j}^{1}\}=\{\bfs_{1}^{1},...,\bfs_{j-1}^{1},\bfs_{j+1}^{1},...,\bfs_{l}^{1}\}$ for $1\leq j\leq l$. The netgain of the $j$th pixel can be rewritten as
\benrr
NG(\bfs_{j}^{1};\wti{S}_{2})
&=&\cL(S_{1}\backslash \cT_{h, -j}^{1},S_{2}\cup \cT_{h, -j}^{1})-\cL(\wti{S}_{1},\wti{S}_{2}),\\
&=&\Big[\cL(S_{1}\backslash \cT_{h, -j}^{1},S_{2}\cup \cT_{h, -j}^{1})-\cL(S_{1},S_{2})\Big]- \Big[\cL(\wti{S}_{1},\wti{S}_{2})-\cL(S_{1},S_{2})\Big],\\
&=& \cD_{1}-\cD_{2}, \quad \textrm{say}.
\eenrr
Akin to the proof of the previous case, we will show that both $\cD_{1}$ and $\cD_{2}$ can be expressed as the sum of the common netgains, those common netgains can be cancelled out, and hence, $\cD_{1}-\cD_{2}$ can be simplified to the sum of three netgains. Note that $(S_{1}\backslash \cT_{h, -j}^{1}, S_{2}\cup \cT_{h, -j}^{1})$ is consequence of transferring $\cT_{h, -j}^{1}=\{\bfs_{1}^{1},...,\bfs_{j-1}^{1},\bfs_{j+1}^{1},...,\bfs_{h}^{1}\}$ from $S_{1}$ to $S_{2}$, and hence, $\cD_{1}$ can be rewritten as a sum of $(h-1)$ netgains as follows:
\benrr
\cD_{1}
&=& NG(\bfs_{1}^{1};S_{1})+\sum_{i=2}^{j-1}NG(\bfs_{i}^{1};S_{1}\backslash \cT_{i-1}^{1})\\
&&+NG(\bfs_{j+1}^{1};S_{1}\backslash \cT_{j-1}^{1})+\sum_{i=j+2}^{h}NG(\bfs_{i}^{1};S_{1}\backslash \cT_{i-1,-j}^{1}).
\eenrr
Given that $\wti{S}_{1}$ can be obtained as a consequence of transferring all pixels of $\cT_{h}^{1}$ with $\bfs_{j}^{1}$ being transferred second to last, i.e., transferring the pixels in the order of $\{\bfs_{1}^{1},...,\bfs_{j-1}^{1},\bfs_{j+1}^{1},...,\bfs_{h-1}^{1},\bfs_{j}^{1},\bfs_{h}^{1}\}$, $\cD_{2}$ can be written as
\benrr
\cD_{2}
&=& NG(\bfs_{1}^{1};S_{1})+\sum_{i=2}^{j-1}NG(\bfs_{i}^{1};S_{1}\backslash \cT_{i-1}^{1})\\
&&+NG(\bfs_{j+1}^{1};S_{1}\backslash \cT_{j-1}^{1})+\sum_{i=j+2}^{h-1}NG(\bfs_{i}^{1};S_{1}\backslash \cT_{i-1,-j}^{1})\\
&&+ NG(\bfs_{j}^{1};S_{1}\backslash \cT_{h-1,-j}^{1})+ NG(\bfs_{h}^{1};S_{1}\backslash \cT_{h-1}^{1}).
\eenrr
Consequently,
\benrr
NG(\bfs_{j}^{1};\wti{S}_{2}) &=& [NG(\bfs_{h}^{1};S_{1}\backslash \cT_{h-1,-j}^{1})-NG(\bfs_{j}^{1};S_{1}\backslash \cT_{h-1,-j}^{1})] - NG(\bfs_{h}^{1};S_{1}\backslash \cT_{h-1}^{1}),\\
&=& [NG(\bfs_{h}^{1};S_{1}\backslash \cT_{h-1,-j}^{1})-NG(\bfs_{j}^{1};S_{1}\backslash \cT_{h-1,-j}^{1})] + NG(\bfs_{h}^{1};\wti{S}_{2}),\\
&>&0\qquad \textrm{in probability},
\eenrr
where the last inequality follows from Lemma \r{cor1} and the fact that $NG(\bfs_{h}^{1};\wti{S}_{2})>0$. $NG(\bfs_{1}^{1};\wti{S}_{2})>0$ in probability can also be shown by transferring $\bfs_{1}^{1}$ second to last when computing $\cD_{2}$, which completes the proof of the lemma.\\
\noi\\
Based on Lemma \r{lem:TS2}, we examine elements originally belonging to $S_{2}$ only when constructing $\cT(\wti{S}_{2})$, which implies $\cT(\wti{S}_{2})=\cT(S_{2})$; next, we transfer all pixels of $\cT(S_{2})$ from $\wti{S}_{2}$ to $\wti{S}_{1}$. Finally, we have $(S_{1}\backslash\cT(S_{1}))\cup \cT(S_{2})$ and $(S_{2}\cup\cT(S_{1}))\backslash\cT(S_{2})$ at the end of the stage; update $S_{1}$ and $S_{2}$ with these two sets for the next stage. Then we repeat this procedure until we get $\cT(S_{1})=\cT(S_{2})=\emptyset$, i.e., there is no need to relocate elements between $S_{1}$ and $S_{2}$ to decrease the distance function. The proposed algorithm is summarized below.
\begin{table}[h!]
\centering
\begin{tabular}{l}
  \hline
  \textbf{The proposed algorithm: } \\
  \hline
  Choose a random initial pair $(S_{1}, S_{2})$ \\
  Set $\cT_{1}=\cT_{2}=\emptyset$\\
  \,\,while $\cT_{1}\neq \emptyset$ or $\cT_{2}\neq \emptyset$\\
  \qquad compute $\cT_{1}:=\cT(S_{1})$ and $\cT_{2}:=\cT(S_{2})$. \\
  \qquad transfer all pixels of $\cT_{1}$ from $S_{1}$ to $S_{2}$.\\
  \qquad transfer all pixels of $\cT_{2}$ from $S_{2}$ to $S_{1}$.\\
  \qquad update $S_{1}$ and $S_{2}$ to $(S_{1}\backslash\cT_{1})\cup \cT_{2}$ and $(S_{2}\backslash\cT_{2})\cup \cT_{1}$, respectively\\
  \,\,end while\\
  Return $S_{1}$ and $S_{2}$\\
  \hline
\end{tabular}
\end{table}

Let $\Omega:=\{A: A\subset S\}$ denote a collection of all subsets of $S$. We shall define a metric to measure a distance between any elements of $\Omega$. For $A\in \Omega$, let $|A|$ denote its cardinality. For real numbers $a,b\in \R$, let $a\vee b:=\max(a,b)$. Define a function  $\del:\Omega\times \Omega\rightarrow \mathbb{N}$ as follows:
\benn
\del(A, B)=\left\{
          \begin{array}{ll}
           \big |\,|A|-|B|\,\big|, & \hbox{if $A\subset B$ or $B\subset A$;} \\
            |A|\vee|B|, & \hbox{if $A\cap B=\emptyset$;}\\
            (|A|- |A\cap B|)\vee(|B|- |A\cap B|), & \hbox{otherwise.}
          \end{array}
        \right.
\eenn
\noi
From the fact that $\big|\,|A|-|B|\,\big|\leq (|A|- |A\cap B|)\vee(|B|- |A\cap B|)\leq |A|\vee|B|$ for all $A,B\in \Omega$, we can see that; (i) a distance between $A$ and $B$ gets smaller as the two sets share more in common; and (ii) $\big|\,|A|-|B|\,\big |$ and $|A|\vee|B|$ play roles of lower and upper bounds for $\del(A,B)$.
\begin{lemma}
$\del$ is a valid metric, that is, for $A,B,C\in \Omega$, the following hold:
\begin{enumerate}
  \item $\del(A,B)=0\Leftrightarrow A=B$,
  \item $\del(A,B)=\del(B,A)$,
  \item $\del(A,B)\leq \del(A,C)+\del(C,B)$.
\end{enumerate}
\end{lemma}
\noi
\textbf{Proof}. Proofs of the first and second claims are trivial. For the last claim, assume $|A|\geq |B|$, i.e., $|A|\vee |B|=|A|$. Therefore, it suffices to show that
\benn
\del(A,C)+ \del(B,C)\geq |A|-|A\cap B|,
\eenn
for the following cases; (i) $B\subset A$; (ii) $A\cap B=\emptyset$; and (iii) $A\cap B\neq \emptyset$ but $B$ is not a subset of $A$. Proofs for the first two cases are straightforward, and hence, we consider the last case only.

Observe that
\benrr
\del(A,C) + \del(C,B) &=&\left\{
     \begin{array}{ll}
       2|C|-(|A\cap C|+|B\cap C|), & \hbox{if $|C|>|A|\geq |B|$;} \\
       |A|+|C|-(|A\cap C|+|B\cap C|), & \hbox{if $|A|\geq |C|> |B|$;} \\
       |A|+|B|-(|A\cap C|+|B\cap C|), & \hbox{if $|A|\geq |B|\geq |C|$.}
     \end{array}
   \right.
\eenrr
To complete the proof of the lemma, it will, therefore, suffice to show the following:
\benrr
&&|A\cap C|+|B\cap C|\leq |C|+ |A\cap B|\quad\textrm{if } |C|> |B|,\\
&&|A\cap C|+|B\cap C|\leq |B|+ |A\cap B|\quad\textrm{if } |C|\leq |B|,
\eenrr
or equivalently,
\benn
|A\cap C|+|B\cap C|\leq |B|\vee |C| + |A\cap B|.
\eenn
Noticing $(A\cap C)\cup(B\cap C) = C\cap (A\cup B)$ gives now
\benn
|A\cap C|+|B\cap C| = |C\cap (A\cup B)| + |A\cap B\cap C|.
\eenn
Hence,
\benrr
|A\cap C|+|B\cap C|&=& |C\cap (A\cup B)| + |A\cap B\cap C|,\\
&\leq& |C|+ |A\cap B|,\\
&\leq& |B|\vee|C|+ |A\cap B|,
\eenrr
where the first inequality follows from $C\cap (A\cup B)\subset C$ and $A\cap B\cap C \subset A\cap B$, thereby completing the proof of the lemma.\\
\noi
With the metric $\del$, we can define a neighborhood of a given $A\in \Omega$:
\benn
\cN_{\xi}(A) = \{B\in \Omega: \del(A,B)\leq \xi \},\qquad \xi>0.
\eenn
Now we are ready to state the main result of this article: the proposed method provides at least a locally optimal solution in probability.
\begin{theorem}
Let $(S_{1}^{Pr}, S_{2}^{Pr})$ denote a solution obtained by the proposed method. Then, there exists a
 neighborhood where $(S_{1}^{Pr}, S_{2}^{Pr})$ is at least an optimal solution in probability, i.e., for all $\epsilon>0$, there exists a $0<\xi<\iny$ such that
\benn
P\Big[\cL(S_{1}^{Pr}, S_{2}^{Pr})-\cL(S_{1}^{*}, S_{2}^{*})\leq 0\Big]\geq 1-\epsilon, \quad \forall S_{1}^{*}\in \cN_{\xi}(S_{1}^{Pr}),\,\, \forall n\geq N_{\epsilon,\xi},
\eenn
where $S_{2}^{*}:=S\backslash S_{1}^{*}$.
\end{theorem}
\noi
\textbf{Proof}. It suffices to show that the claim holds for $\xi=1$. To conserve a space, let $\cL^{*}$ and $\cL^{Pr}$ denote $\cL(S_{1}^{*}, S_{2}^{*})$ and $\cL(S_{1}^{Pr}, S_{2}^{Pr})$, respectively. When $S_{1}^{*}\in \cN_{\xi}(S_{1}^{Pr})$, one of the following is true; (i) $S_{1}^{*}\subset S_{1}^{Pr}$;  (ii) $S_{1}^{*}\supset S_{1}^{Pr}$; and (iii) neither of (i) nor (ii) is true. To begin with, consider the first case. Since $\xi=1$ and $S_{1}^{*}\subset S_{1}^{Pr}$, $S_{1}^{*}$ contains all elements of $S_{1}^{Pr}$ but one element, say $\mbf{w}_{1}$: $S_{1}^{Pr}=S_{1}^{*}\cup \{\mbf{w}_{1}\}$. Subsequently, we have $\cL^{*} - \cL^{Pr}=NG(\mbf{w}_{1};S_{1}^{Pr})$. Note that $NG(\mbf{w}_{1};S_{1}^{Pr})$ should be greater than or equal to 0; otherwise, $\mbf{w}_{1}\in \cT(S_{1}^{Pr})$, which implies relocation of $\mbf{w}_{1}$ will decrease the distance function , and hence,
\benn
\cL(S_{1}^{Pr}\backslash\{\mbf{w}_{1}\}, S_{2}^{Pr}\cup \{\mbf{w}_{1}\})<\cL(S_{1}^{Pr}, S_{2}^{Pr}),
\eenn
thereby contradicting $(S_{1}^{Pr}, S_{2}^{Pr})$ is the optimal solution.

For the case that $S_{1}^{*}\supset S_{1}^{Pr}$, the same argument can be applied; only difference between the first and second cases is we transfer an element from $S_{2}^{Pr}$ to $S_{1}^{Pr}$, and hence, we replace $S_{1}^{Pr}$ with $S_{2}^{Pr}$ in the above argument.

Finally, consider the last case. $S_{1}^{*}\in \cN_{\xi}(S_{1}^{Pr})$ with $\xi=1$ implies that there are pixels $\mbf{w}_{1}$ and $\mbf{v}_{1}$ such that $S_{1}^{Pr}\backslash S_{1}^{*}=\{\mbf{w}_{1}\}$ and $S_{1}^{*}\backslash S_{1}^{Pr}=\{\mbf{v}_{1}\}$. Let $A_{1}$ and $A_{2}$ denote $S_{1}^{Pr}\cap S_{1}^{*}\neq \emptyset$ and $A_{1}^{c}$, respectively, which trivially implies $A_{1}\in \cN_{\xi}(S_{1}^{Pr})$ and
\ben\lel{eq:S_and_A}
S_{1}^{Pr}=A_{1}\cup \{\mbf{w}_{1}\}\textrm{ and } S_{1}^{*}=A_{1}\cup \{\mbf{v}_{1}\}.
\een
As in the first case, $A_{1}\subset S_{1}^{Pr}$ implies
\ben\lel{eq:Pr-A}
\cL(S_{1}^{Pr},S_{2}^{Pr})-\cL(A_{1},A_{2})\leq 0.
\een
From (\r{eq:S_and_A}) and $A_{2}=A_{1}^{c}$, it is not difficult to see that $A_{2}=S_{2}^{Pr}\cup \{\mbf{w_{1}}\}$, i.e., $S_{2}^{Pr}=A_{2}^{(-\{\mbf{w_{1}}\})}$. Similarly, $A_{2}=S_{2}^{*}\cup \{\mbf{v_{1}}\}$, and hence, $S_{2}^{*}=A_{2}^{(-\{\mbf{v_{1}}\})}$.
For $\mbf{v_{1}}\in S_{2}^{Pr}$,
\benr\lel{eq:sgn}
\textrm{sgn}\Big[NG(\mbf{v_{1}};S_{2}^{Pr})\Big]
&=&\textrm{sgn}\Big[NG(\mbf{v_{1}};A_{2}^{(-\{\mbf{w_{1}}\})})\Big],\nonumber\\
&=& \textrm{sgn}\Big[NG(\mbf{v_{1}};A_{2})\Big] \quad \textrm{ in probability},
\eenr
where the last equality follows from Lemma \r{eq:NG_positive}. Since $(S_{1}^{Pr}, S_{2}^{Pr})$ is the optimal solution, we should have
\ben\lel{eq:NG_v1}
NG(\mbf{v_{1}};S_{2}^{Pr}) = \cL(S_{1}^{Pr}\cup \{\mbf{v_{1}}\}, S_{2}^{Pr}\backslash \{\mbf{v_{1}}\})-\cL(S_{1}^{Pr}, S_{2}^{Pr})\geq 0.
\een
Consequently,
\benrr
\cL(S_{1}^{*}, S_{2}^{*})-\cL(A_{1}, A_{2})
&=&  \cL(A_{1}\cup \{\mbf{v_{1}}\}, A_{2}\backslash \{\mbf{v_{1}}\})-\cL(A_{1}, A_{2}),\\
&=& NG(\mbf{v_{1}};A_{2}),\\
&\geq&0\quad \textrm{ in probability},
\eenrr
where the last inequality follows from (\r{eq:sgn}) and (\r{eq:NG_v1}). Finally, the last inequality together with (\r{eq:Pr-A}) enables one to conclude
\benn
\cL(S_{1}^{Pr},S_{2}^{Pr})\leq\cL(S_{1}^{P*},S_{2}^{P*}) \quad \textrm{in probability,}
\eenn
thereby completing the proof of the theorem.\\
\noi\\
The next paragraph describes how the distance function behaves on the given domain in the case of $K=2$. 
Let $S_{1}^{\dagger}=\{\mbf{s}_{1}^{1},\mbf{s}_{2}^{1},...,\mbf{s}_{n_{1}}^{1}\}$ and $S_{2}^{\dagger}=\{\mbf{s}_{1}^{2},\mbf{s}_{2}^{2},...,\mbf{s}_{n_{2}}^{2}\}$. Note that $S_{2}^{\dagger}=S\backslash S_{1}^{\dagger}$. Define
$T_{1}^{1}:= S_{1}^{(+n_{2})} = S_{1}^{\dagger}\cup\{\bfs_{n_{2}}^{2}\}$, i.e., the last element of $S_{2}^{\dagger}$ will be transferred to $S_{1}^{\dagger}$. Next, let $T_{2}^{1}:=S\backslash T_{1}^{1}$ denote the complement of $T_{1}^{1}$. Therefore, the cardinalities of $T_{1}^{1}$ and $T_{2}^{1}$ are $n_{1}+1$ and $n_{2}-1$, respectively. Recursively, define
\benrr
T_{1}^{2} &:=& T_{1}^{1,(+(n_{2}-1))} = S_{1}^{\dagger}\cup\{\bfs_{n_{2}-1}^{2}, \bfs_{n_{2}}^{2}\},\\
T_{2}^{2} &:=& S-T_{1}^{2} =  S_{2}^{\dagger}-\{\bfs_{n_{2}-1}^{2}, \bfs_{n_{2}}^{2}\},\\
\vdots & & \vdots \\
T_{1}^{n_{2}} &:=& T_{1}^{(n_{2}-1),(+1)} = S_{1}^{\dagger}\cup\{\bfs_{1}^{2},...,\bfs_{n_{2}-1}^{2}, \bfs_{n_{2}}^{2}\},\\
T_{2}^{n_{2}} &:=& S-T_{1}^{2} =  S_{2}^{\dagger}-\{\bfs_{1}^{2},...,\bfs_{n_{2}-1}^{2}, \bfs_{n_{2}}^{2}\}.
\eenrr
Observe that the superscript denotes the ordinal number of stage and the number of all elements transferred from $S_{2}^{\dagger}$ to $S_{1}^{\dagger}$ until the stage. It is plain to see that $T_{1}^{n_{2}}\equiv S$ and $T_{2}^{n_{2}}\equiv \emptyset$. Next, we define a sequence of sets in the opposite manner
\benrr
R_{1}^{1}&:=& S_{1}^{(-n_{1})} = S_{1}^{\dagger}-\{\bfs_{n_{1}}^{1}\},\\
R_{2}^{1}&:=& S_{2}^{(+n_{1})} = S_{2}^{\dagger}\cup\{\bfs_{n_{1}}^{1}\},\\
R_{1}^{2} &:=& R_{1}^{1,(-(n_{1}-1))} = S_{1}^{\dagger}-\{\bfs_{n_{1}-1}^{1}, \bfs_{n_{1}}^{1}\},\\
R_{2}^{2} &:=& S-R_{1}^{2} =  S_{2}^{\dagger}\cup\{\bfs_{n_{1}-1}^{1}, \bfs_{n_{1}}^{1}\},\\
\vdots & & \vdots \\
R_{1}^{n_{1}} &:=& R_{1}^{(n_{1}-1),(-1)} = S_{1}^{\dagger}-\{\bfs_{1}^{1},...,\bfs_{n_{1}-1}^{1}, \bfs_{n_{1}}^{1}\},\\
R_{2}^{n_{1}} &:=& S-R_{1}^{2} =  S_{2}^{\dagger}-\{\bfs_{1}^{1},...,\bfs_{n_{1}-1}^{1}, \bfs_{n_{1}}^{1}\},
\eenrr
where $R_{1}^{n_{1}}=\emptyset$ and $R_{2}^{n_{1}}=S$. Finally, we have
\benn
\emptyset = R_{1}^{n_{1}}\subset\cdots\subset R_{1}^{1}\subset S_{1}^{\dagger}\subset T_{1}^{1}\subset \cdots\subset T_{1}^{n_{2}}=S,
\eenn
and
\benn
S = R_{2}^{n_{1}}\supset\cdots\supset R_{2}^{1}\supset S_{2}^{\dagger}\supset T_{2}^{1}\supset \cdots\supset T_{2}^{n_{2}}=\emptyset.
\eenn
Define a collection of pairs of sets as follows
\benn
\boldsymbol{\cT} := \{(R_{1}^{n_{1}}, R_{2}^{n_{1}}),...,(R_{1}^{1}, R_{2}^{1}), (S_{1}^{\dagger}, S_{2}^{\dagger}),(T_{1}^{1}, T_{2}^{1}),... (T_{1}^{n_{2}}, T_{2}^{n_{2}})\}.
\eenn
Note that $\boldsymbol{\cT}\subset \boldsymbol{\cS}^{2}$. \\
\begin{figure}[h]
\centering
\includegraphics[width=0.75\textwidth]{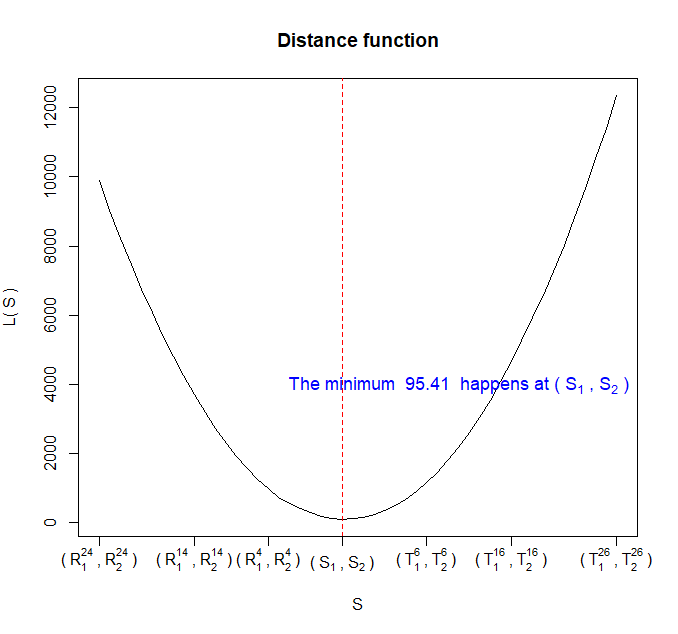}
\caption{A graph of $\cL$ over $\boldsymbol{\cT}$.}\lel{Fig:graph_runtime}
\end{figure}
\noi\\
\textbf{Example 2.} Recall the image from Example 1 where $S_{1}^{\dagger}$ and $S_{2}^{\dagger}$ denote sets of white and black pixels, respectively. Without noise, the segmentation of $S_{1}^{\dagger}$ is not challenging at all. Even though the noise in this example is not strong enough to make the segmentation challenging, the presence of the noise still adds more difficulty than would otherwise be the case. With the real observed image with noise, the segmentation of the white circle amounts to estimating $S_{1}^{\dagger}$ by searching the minimum of the distance function $\cL$. Figure \r{Fig:graph_runtime} shows a graph of the distance function over $\boldsymbol{\cT}$. As displayed in the figure, $\cL$ attains the minimum at $(S_{1}^{\dagger}, S_{2}^{\dagger})$, the true sets of the white and black pixels. This result also closely accords with the argument in Remark \r{rem:penalty}. Here, the optimal solution completely overlaps with $(S_{1}^{\dagger}, S_{2}^{\dagger})$ due to very weak noise; otherwise, there will be considerable disagreement between them.

\section{Simulation studies}
\subsection{General setup}
Recall the model (\r{eq:Model}). Through this section, we use a collection of pixels
\benn
S=\{(i,j): 1\leq i\leq 200, 1\leq j\leq 200\},
\eenn
i.e., $S$ is a $200\times 200$ square. As in Example 1, we consider $K=2$ and assume that $g:\R^{2}\rightarrow\R$ takes $p_{1}=1$ and $p_{2}=0$ over $S_{1}^{\dagger}$ and $S_{2}^{\dagger}$, respectively. For the error (or noise) in the model (\r{eq:Model}), we assume that it follows a normal distribution with a mean of zero and a standard deviation of $\sigma$. Kim \cite{KimJ2018}, \cite{KimJ2020} showed that the MD estimators of linear models with both independent and dependent errors -- which follow a wide range of distributions, including a normal, logistic, Cauchy, and  mixture of two distributions -- perform well. Similar to his finding, the MD estimators of this study show similar performance regardless of distributions of the error, and hence, we report the simulation result corresponding to the normal error only. When generating the normal error, we try $\si=$0.1, 0.5, and 0.8 for the comparison purpose: the corresponding errors are referred to as mild, moderate, and severe errors, respectively. For the computational simulation in this study, RStudio 1.1.463 is used; the CPU used for gauging the computational speed is Intel(R) Core(TM) i7-10700.

The size of all images used for the simulation studies is $200\times 200$. A segmentation task for a simulated $200\times 200$ image -- e.g., the segmentation of the white circle as in Example 1 -- is not computationally expensive in that it does not take much time. However, some other real images -- e.g., magnetic resonance images --  of a large dimension will take substantial amount of time. Therefore, it is absolutely imperative to address the high computational cost ensuing from a segmentation task for a larger image. To make a breakthrough in reducing the cost, we will combine the proposed method with a well-known method in the next section: the patch-wise segmentation.

\subsection{Patch-wise segmentation}
To establish the superiority of the patch-wise segmentation over the usual segmentation without using patch images, we will demonstrate that the former completes a given segmentation task much faster. For an input image used for segmentation, we first create an image of a cross without any noise: see, e.g., Figure \r{Fig:Wo_noise}. Next, randomly generated noise $\vep_{i}$ from $N(0,\,0.5^{2})$ will be added to the original image, and the final image will be like the third one of Figure \r{Fig:noise_2}. To assess how the patch size affects the computational time of the segmentation, various square patches whose lengths range between 4 and 100 will be tried. For each square patch, we will repeat the segmentation 100 times and record the average computational time of the 100 trials. Table \r{tbl:runtime} reports the results of the patch-wise segmentation with various patch sizes being tried.
\begin{table}[h]
\centering
\begin{tabular}{c c c c } 
\hline
 $L$  & $T$ & $N$ & $T/N$   \\
 (length) & (seconds) & (\# of patches) & ($\times 10^{-5}$) \\
\hline
 4    &0.725   &9,801 &7.406   \\
 8    &0.993   &9,409 &10.563  \\
 16   &1.652   &8,649 &19.109  \\
 32   &5.830   &7,225 &80.703   \\
 64   &78.790  &4,761 &165.492 \\
 80   &192.128 &3,721 &516.334 \\
 100  &485.554 &2,601 &1,866.797\\
\hline
\end{tabular}
\caption{Computational times when patch images of various sizes are used.}\lel{tbl:runtime}
\end{table}

The first column denoted by $L$ represents the length of the square patch used for the segmentation while the second column denoted by $T$ reports the average computational time when the square patch of the corresponding length is used. For example, using a square patch of length 32 requires 5.830 seconds for the segmentation of the entire $200\times 200$ input image while only 0.725 second is taken for the same segmentation when a square patch of length 4 is used. The third column denoted by $N$ represents the number of extracted patch images during the entire segmentation. Since a stride of 2 is used when a square patch slides both horizontally and vertically, $N$ is equivalent to $[(200-L)/2+1]^{2}$. Following the third columns, the fourth column ($T/N$) reports the average time taken for segmentation of a single square patch where the figure in the parenthesis is a unit of computational time. For example, it will take $10.563\times 10^{-5}$ $(=0.993/9,409)$ seconds on average for the segmentation of an $8\times 8$ square patch.

It is plain beyond misapprehension that the computational time decreases dramatically as a patch of a smaller size is used. As reported in the table, using a $4\times 4$ patch yields the least amount of computational time. Motivated by this fact, the patch-wise segmentation with the $4\times 4$ patch will be used for the following analysis unless specified otherwise.

\subsection{Image segmentation of simulated images}
In this section, we will try various simulated images (circle, square, triangle, and star) with noise for segmentation. To visualize the performance of the proposed method, we invert the colors of the resulting images after segmentation, i.e., transforming white pixels to black pixels or vice versa. Figures \r{Fig:Wo_noise}-\r{Fig:noise_3} show original images together with their segmented outputs: Figure \r{Fig:Wo_noise} shows the result pertaining to the original image without any noise while Figures \r{Fig:noise_1}, \r{Fig:noise_2}, and \r{Fig:noise_3} show the results when the original image is contaminated with mild, moderate, and severe noise, respectively.
\ifnum \value{show}=1{
\begin{figure}[h]
\centering
\includegraphics[width=0.95\textwidth]{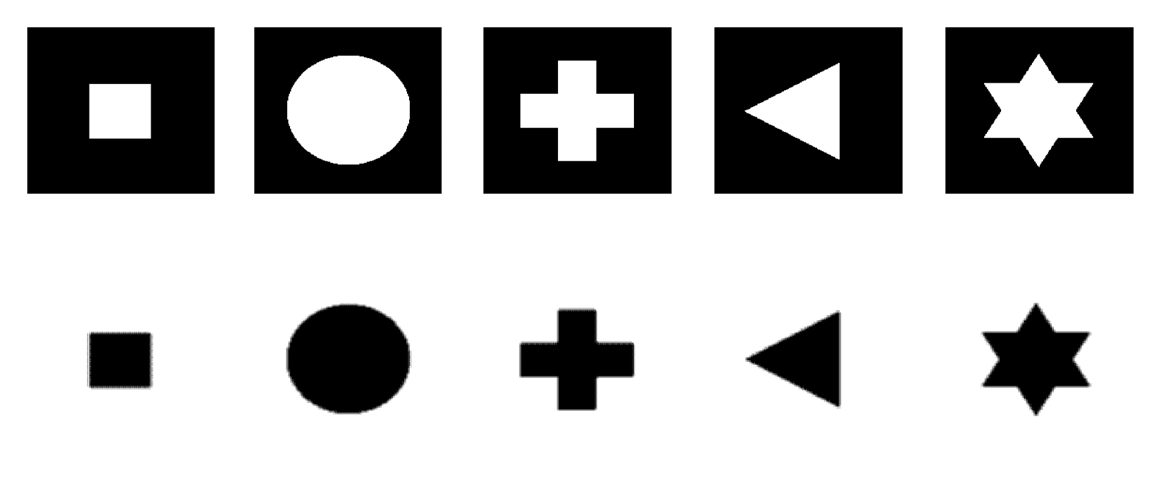}
\caption{Original images (top) and segmented images (bottom)}\lel{Fig:Wo_noise}
\end{figure}
}\fi

\ifnum \value{show}=1{
\begin{figure}[h]
\centering
\includegraphics[width=0.95\textwidth]{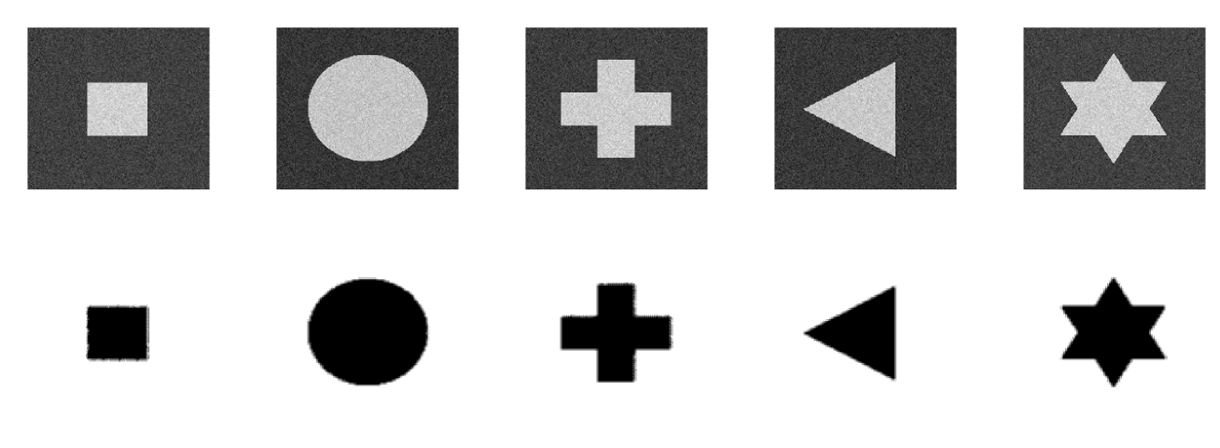}
\caption{Original images with mild errors (top) and segmented images (bottom)}\lel{Fig:noise_1}
\end{figure}
}\fi

\ifnum \value{show}=1{
\begin{figure}[h]
\centering
\includegraphics[width=0.95\textwidth]{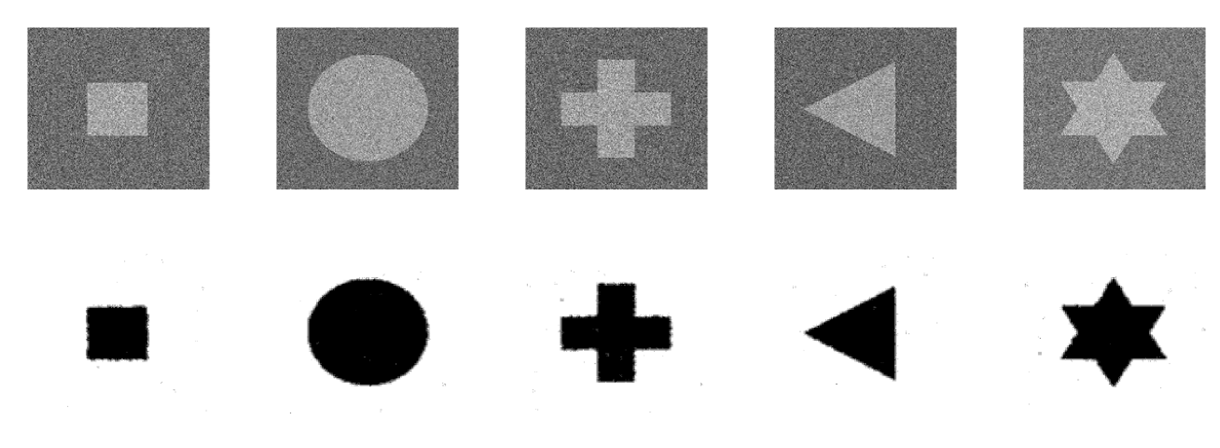}
\caption{Original images with moderate errors (top) and segmented images (bottom).}\lel{Fig:noise_2}
\end{figure}
}\fi

\ifnum \value{show}=1{
\begin{figure}[h]
\centering
\includegraphics[width=0.95\textwidth]{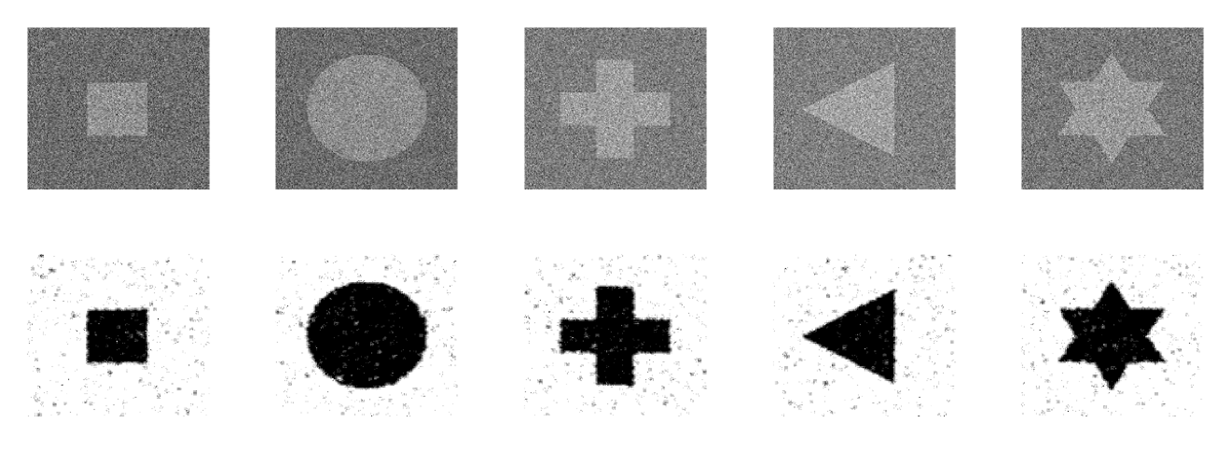}
\caption{Original images with severe errors (top) and segmented images (bottom).}\lel{Fig:noise_3}
\end{figure}
}\fi
Several points are worth mentioning here. First, the proposed method returns perfectly-segmented images when there is no noise (Figure \r{Fig:Wo_noise}) or exist mild noise (Figure \r{Fig:noise_1}). Even in the presence of moderate noise, the proposed method shows very good segmentation performance; there are only a few false-positively segmented (FPS) pixels which are wrongly segmented as white pixels. Note that the number of FPS pixels in the final segmented image increases when images contain severe noise as shown in Figure \r{Fig:noise_3}. Second, it is clear to see the border lines between $S_{1}^{\dagger}$ (white) and $S_{2}^{\dagger}$ (black) get blurred as the noise gets stronger and poses a serious impediment to accurate segmentation. A closer look at the images with the severe noise reveals that the border lines of the segmented images are not straight but all saw-edged. In an effort to reduce (or remove completely if possible) those FPS pixels, we integrate the patch-wise segmentation with another well-celebrated digital filtering technique: median filtering. The median filtering has been popular and widely used in digital image processing for its several merits: see, e.g., Huang et.\,al \cite{Huang1979} for more details. Figure \r{Fig:Median_Filtering} shows the results of the patch-wise segmentation with or without the median filtering. As shown in the figure, it is crystal-clear that there is a stark difference between the two figures: most of the FPS pixels are removed when the median filtering is applied. Thus, the median filtering will be also used together with the proposed method unless otherwise noted.
\ifnum \value{show}=1{
\begin{figure}[h!]
    \centering
    \begin{subfigure}[b]{0.45\textwidth}
        \includegraphics[height=60mm, width=60mm]{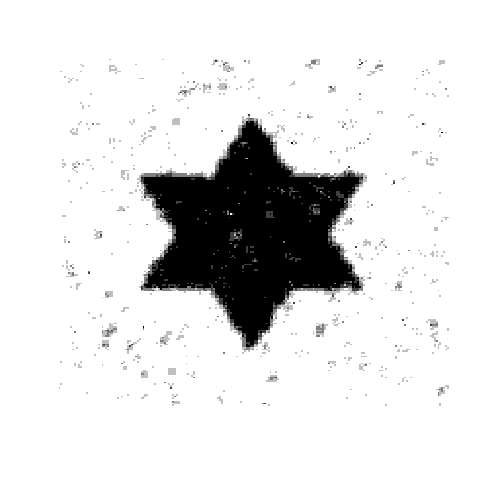}
    \end{subfigure}
    \,\,
    \begin{subfigure}[b]{0.45\textwidth}
        \includegraphics[height=60mm, width=60mm]{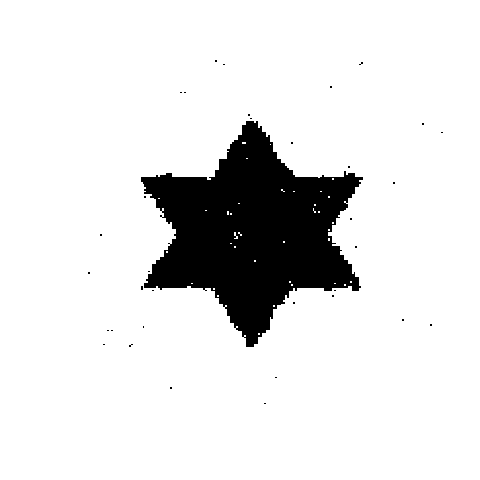}
    \end{subfigure}
    \caption{Segmented image without (left) and with (right) the median filtering.}\lel{Fig:Median_Filtering}
\end{figure}
}\fi
\noindent\\

As shown in the previous figures, the proposed method performs the image segmentation well regardless of the presence of noises. However, this result is confined to simple images; the complexity of usual real images (a cat, dog, etc) alters the case, and therefore, the previous result is not promising. To substantiate that the proposed method has a potential real-world application, we, therefore, should try a more complex image: a $200\times 200$ pseudo-QR code image that resembles a QR code. For generating the image, we obtain 20,000 pairs of $(i,j)$ by randomly generating $i$ and $j$ from the discrete uniform distribution on $\{1,2,...,200\}$ and assign 1 to the $(i,j)$th entry of the image for the pixel value while 0 is assigned to the rest entries of the image. Therefore, the resulting image will contain the same number of white and black pixels. Figure \r{Fig:QR_images} reports the result pertaining to the segmentation of the pseudo-QR code image.
\ifnum \value{show}=1{
\begin{figure}[h!]
\centering
    \begin{subfigure}[b]{0.45\textwidth}
        \includegraphics[height=40mm, width=50mm]{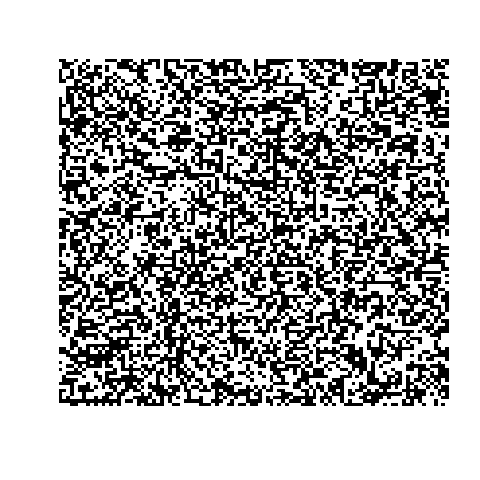}
    \end{subfigure}
    \begin{subfigure}[b]{0.45\textwidth}
        \includegraphics[height=40mm, width=50mm]{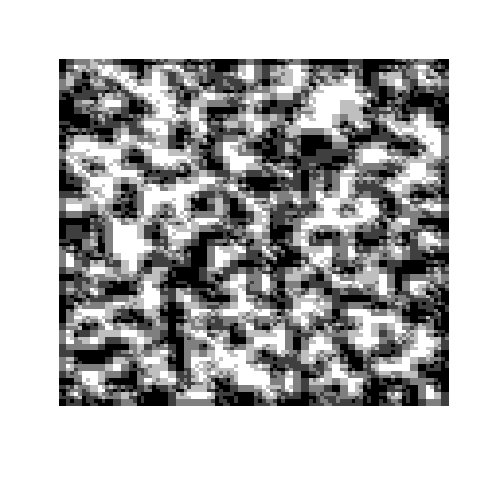}
    \end{subfigure}
    \caption{The original pseudo-QR image (left) and segmented image (right)}\lel{Fig:QR_images}
\end{figure}
}\fi

Even just a quick glance reveals that the performance of the proposed method deteriorates to a large extent when compared with the previous cases of simple images. One point worth noting at this juncture is that the full-fledged noise is not even introduced yet in the image. The competence to handle the presence of noise and complexity of a target image is an indispensable virtue that the proposed method should retain in order to remain a competitive method for image segmentation.

\subsection{Segmenting-together strategy}
As shown in the pseudo-QR image, the proposed method displayed a disappointing performance. To redress this issue, we propose a novel strategy as an addendum to the proposed method: we refer to this strategy as ``segmenting-together strategy" that means, ad litteram, segmenting a group of pixels of similar colors together rather than a disparate group of pixels. Figure \r{Fig:trans} describes the general procedure of the segmenting-together strategy when it is applied to a $4\times 5$ image.
\ifnum \value{show}=1{
\begin{figure}[h!]
\centering
\includegraphics[width=0.7\textwidth]{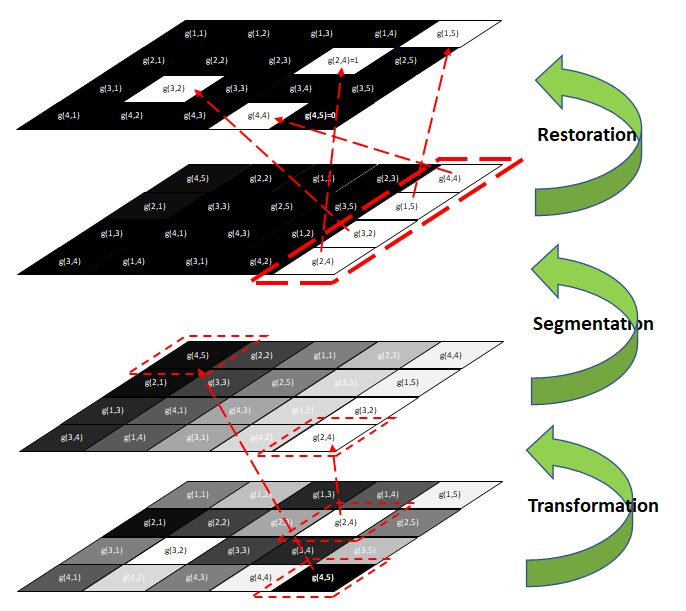}
\caption{Transformation and retrieval of an image through sorting pixels: an original image(bottom), a transformed image(middle) and a retrieved image.}\lel{Fig:trans}
\end{figure}
}\fi
The procedure consists of three stages; (1) transforming the original image; (2) segmenting a group of bright pixels; and (3) restoring the segmented pixels to their original entries. In the stage of transformation, pixels will be sorted and arranged in increasing order of pixel values; the (4,5)th entry, as a case in point, of the original image in Figure \r{Fig:trans} having the least pixel value (=0) will be relocated to the (1,1)th entry of the transformed image while the (2,4)th entry of the original image with the largest pixel value of 1 will be relocated to the (4,5)th entry of the transformed image. For the transformation of an $M\times N$ image, we can define an associated one-to-one mapping $\phi:\mathbb{N}\times \mathbb{N}\rightarrow \mathbb{N}\times \mathbb{N}$
\benn
\phi(i_{2},j_{2}) = (i_{1}, j_{1}), \quad 1\leq i_{1}, i_{2}\leq M,\,\, 1\leq j_{1}, j_{2}\leq N,
\eenn
where $(i_{1},j_{1})$ represents the $(i_{1},j_{1})$th entry of the original image while $(i_{2},j_{2})$ represents the $(i_{2},j_{2})$th entry of the transformed image. Through referring to $\phi$, we can ascertain the original entry of any given entry of the transformed image, and hence, the original image can be retrieved at any time.

In the original image of the figure, let the four brightest pixels -- the (1,5)th, (2,4)th, (3,2)th, and (4,4)th entries of the image -- constitute a region of interest (ROI). After the transformation, these pixels will be relocated to the last column of the transformed image. Upon the completion of transforming the original image, we apply the proposed method to the resulting transformed image for segmentation. Assume that only those four pixels of ROI survive a segmentation process while others do not. Then, we highlight those survived pixels by changing their pixel values to 1 while transforming other pixels completely black by assigning 0 for the pixel value. Finally, those segmented pixels will be restored to their original entries by referring to the mapping $\phi$.

Figure \r{Fig:procedure} compares outcomes obtained from the proposed method only (middle) and the proposed method in conjunction with the segmenting-together strategy (right).
\ifnum \value{show}=1{
\begin{figure}[h!]
\centering
    \begin{subfigure}[b]{0.31\textwidth}
        \includegraphics[height=60mm, width=60mm]{Image/Img/Random}
    \end{subfigure}
    \begin{subfigure}[b]{0.31\textwidth}
        \includegraphics[height=60mm, width=60mm]{Image/Img/Random_Seg}
    \end{subfigure}
    \begin{subfigure}[b]{0.31\textwidth}
        \includegraphics[height=60mm, width=60mm]{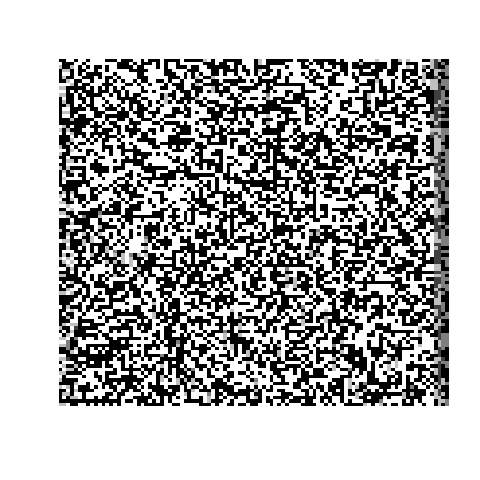}
    \end{subfigure}
    \caption{Original QR images (left), segmented image by the proposed method (middle) and image by the proposed method together with segmenting-together strategy (right).}\lel{Fig:procedure}
\end{figure}
}\fi
From the figure, it is immediately apparent that the proposed method shows a remarkable improvement in performance when rigged with the segmenting-together strategy, thereby proving the preponderant role of the strategy. Its significance becomes clearer when we investigate the outcome obtained from the proposed method with the strategy in greater detail. To sustain the last argument, we invert the segmented image, superimpose it on the original image, and examine how much the overlaid image fills the original image. If the segmentation is perfect, then the resulting image will be completely black. On the contrary, the resulting image will be the original image itself if the segmentation goes completely awry; otherwise, the resulting image will range from the original image to a completely black image, pari passu to the performance of the proposed method. Figure \r{Fig:together} reports the result of the overlay analysis.
\ifnum \value{show}=1{
\begin{figure}[h!]
\centering
\includegraphics[width=0.95\textwidth]{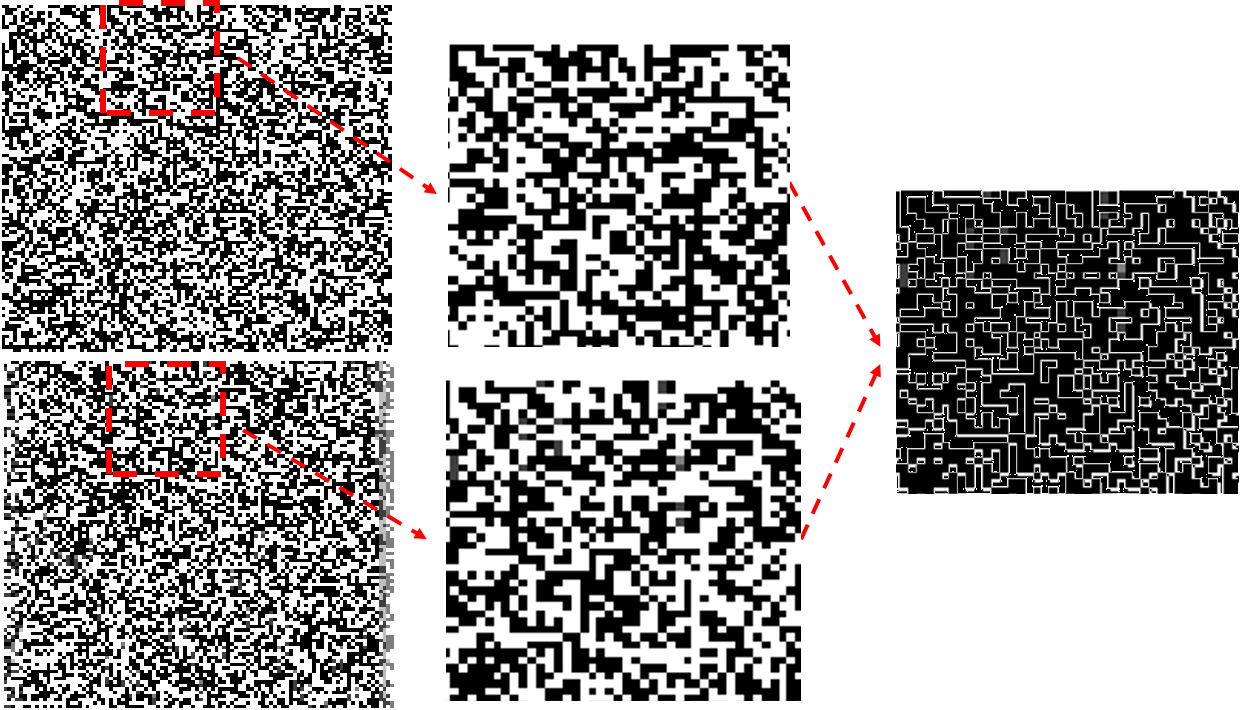}
\caption{Comparison of the original image and the resulting image from the segmenting-together-strategy.}\lel{Fig:together}
\end{figure}
}\fi
As reported in the figure, the overlay of the inverted image after segmentation almost perfectly fills the original image, and hence, the resulting image is almost black. To numerically gauge the exquisite performance of the proposed method with the strategy and demonstrate its superiority, we employ another measure: Dice similarity coefficient (DSC). For given two images $A$ and $B$, the DSC is defined as
\benn
DSC = \frac{2|A\cap B|}{|A|+|B|},
\eenn
where $|\cdot|$ denotes the number of pixels in a image while the intersection of two images denotes the common, overlapped image between them. The DSC value ranges from 0 to 1, indicating no and complete overlaps, respectively. In the following segmentation of simulated and real images, the DSC is adopted here to validate the proposed method. Note that the validation requires two images as inputs: $A$ and $B$. For simulated images, we use original images before adding noise and the segmented image since the original images are available. In case of the real images, the original images without noise is not retrievable, and hence, we use ``putative" original images that are labeled by experts of those images.

Recall Figure \r{Fig:procedure} where the pseudo-QR image without noise was tried for segmentation.
The DSC value of the segmented image from the proposed method only (middle of the figure) is 0.625, which implies it correctly matches only 62.5\% of the entire original image. After the proposed method is combined with the segmenting-together strategy, the DSC surges to 0.942, thereby showing signs of drastic improvement. This promising result still holds true even in the presence of noise, which is illustrated in Figure \r{Fig:DSC}.
\ifnum \value{show}=1{
\begin{figure}[h!]
\centering
    \begin{subfigure}[b]{0.45\textwidth}
        \includegraphics[height=60mm, width=60mm]{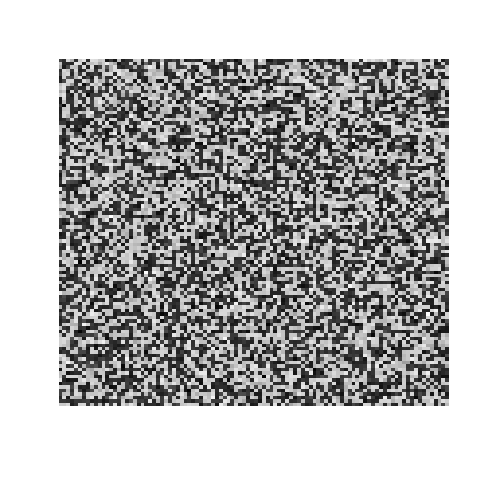}
    \end{subfigure}
    \begin{subfigure}[b]{0.45\textwidth}
        \includegraphics[height=60mm, width=60mm]{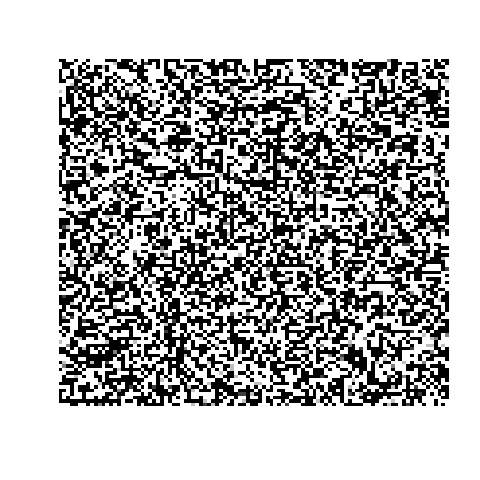}
    \end{subfigure}
    ~
    \begin{subfigure}[b]{0.45\textwidth}
        \includegraphics[height=60mm, width=60mm]{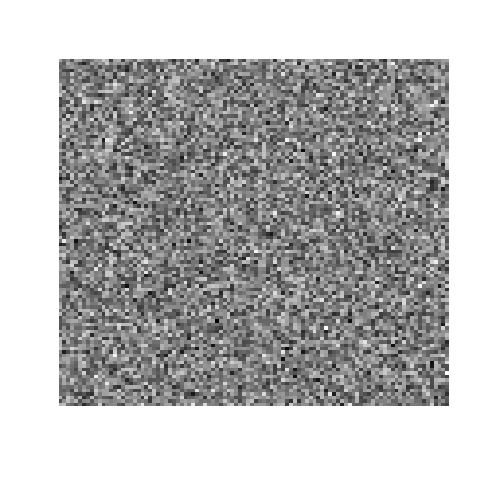}
    \end{subfigure}
    \begin{subfigure}[b]{0.45\textwidth}
        \includegraphics[height=60mm, width=60mm]{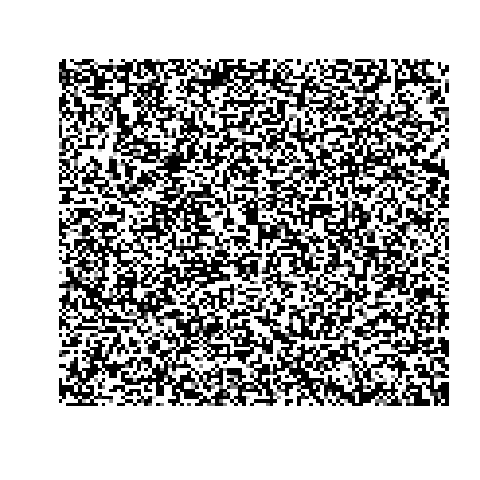}
    \end{subfigure}
    ~
    \begin{subfigure}[b]{0.45\textwidth}
        \includegraphics[height=60mm, width=60mm]{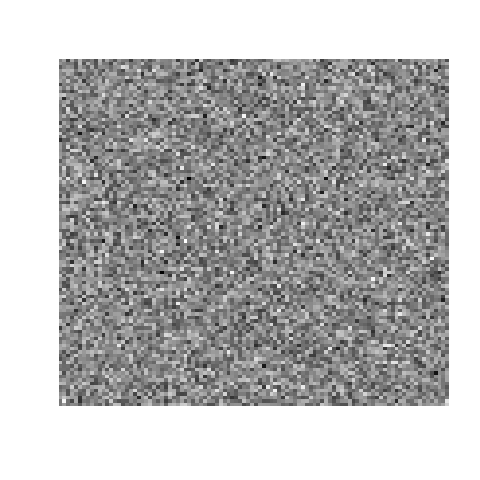}
    \end{subfigure}
    \begin{subfigure}[b]{0.45\textwidth}
        \includegraphics[height=60mm, width=60mm]{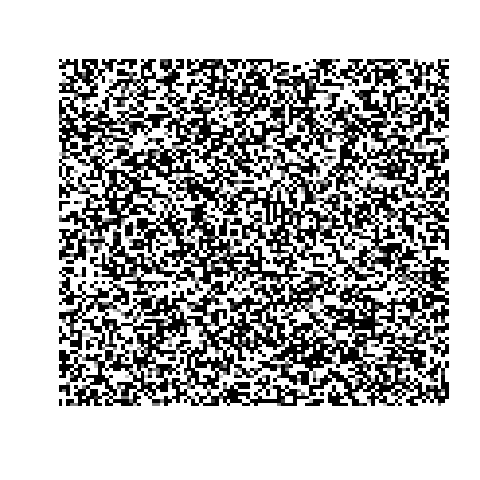}
    \end{subfigure}
    \caption{Original QR images with various errors (left) and segmented images (right).}\lel{Fig:DSC}
\end{figure}
}\fi
Figure \r{Fig:DSC} reports the segmentation results (right) when the QR images (left) contaminated by various noise -- mild (top), moderate (middle), and severe (bottom) -- are segmented through the proposed method with the strategy. Then, we obtain the DSC values of 0.931, 0.798, and 0.707 for the mild, moderate, and severe noise, respectively. Recall the case of the pseudo-QR image without any noise where the proposed method only yielded the DSC value of 0.625. When the proposed method is combined with the strategy, it yields still a better DSC value (0.707) even in the presence of the severe noise.
\subsection{Real examples}
The previous section observed that the felicitous conjunction of the proposed method and the segmenting-together strategy comes as an amazement. However, the results obtained from the simulation studies in the previous section should be treated and interpreted with considerable caution. It is not surprising to frequently observe that the excellence of methods with the simulated data is belied by the poor performance in real data: they start brilliantly with the simulated data but sink into or below mediocrity with real data. At this juncture, we have to answer the following question: can the proposed method replicate the theoretical result obtained so far when it is applied to the real data? Successful replication will lend credence to the proposed method while failure to do that will cast a pall of suspicion over the method. In view of this, showing good performance for real data is crucial. To this end, this section; (1) assesses the performance of the proposed method when the real data is used for segmentation; and (2) demonstrates that the proposed method will remain competitive even when it is adopted for handling real data, thereby consolidating its position as the potential option for the image segmentation.

For the real data, we use the magnetic resonance (MR) images of brain tumors from Baid et al.\,\,\cite{Baid2021}. In the MR images of the brain, the tumors are denoted by bright colors while normal cells are denoted by dark colors. In the gray-scale MR images, tumors have a pixel value of or close to 1 (white) while normal cells have a pixel value of or close to 0 (black). Figures \r{Fig:RSNA1} and \r{Fig:RSNA3} report the result when the proposed method together with the segmenting-together strategy is employed for different types of two MR images: Fluid-attenuated inversion recovery (FLAIR) and T1-weighted images.
\ifnum \value{show}=1{
\begin{figure}[h!]
\centering
\includegraphics[width=0.95\textwidth]{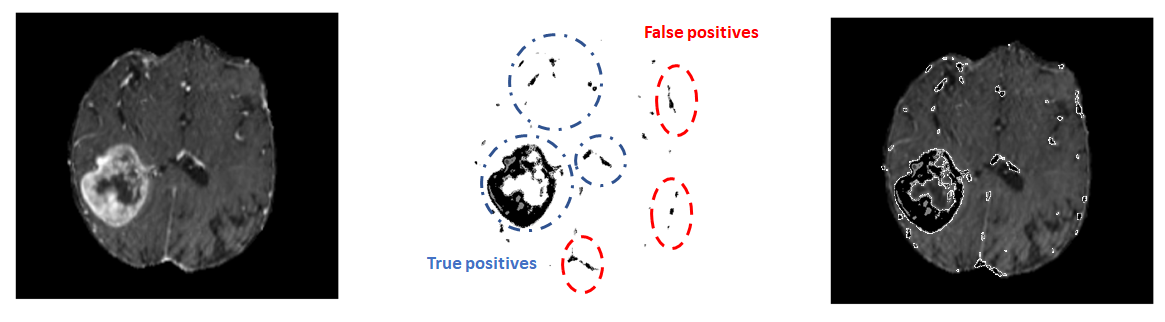}
\caption{A FLAIR image of brain tumors (left), the segmented image by the proposed method (middle), and the overlaid image (right) }\lel{Fig:RSNA1}
\end{figure}
}\fi
Figure \r{Fig:RSNA1} shows the original FLAIR MR image (left), the inverted image of the segmentation (middle), and the overlay of the segmented image over the original one (right). Based on the overlay of the segmented image, the proposed method seems to work properly: it is, at least, not a meretricious method which works for the simulated data only. As shown in the segmented image, there are still some FPS pixels; however, most of all of those are located in the area of the skull which in the original image is denoted by as much bright pixels as tumors. During the preparation of the MR image before the segmentation -- which is called ``preprocessing" -- those bright pixels in the skull are removed. Therefore, those FPS pixels might be imputed to a less careful preprocessing procedure of the MR image. If a better-preprocessed image were used, then those types of FPS pixels would disappear. One promising fact here is that the proposed method successfully detects tumors of a very small size which are located inside of the top blue circle in the middle figure: we surmise that this is originated from the segmenting-together strategy. Without it, those small tumors could have not been detected. 
\ifnum \value{show}=1{
\begin{figure}[h!]
\centering
\includegraphics[width=0.95\textwidth]{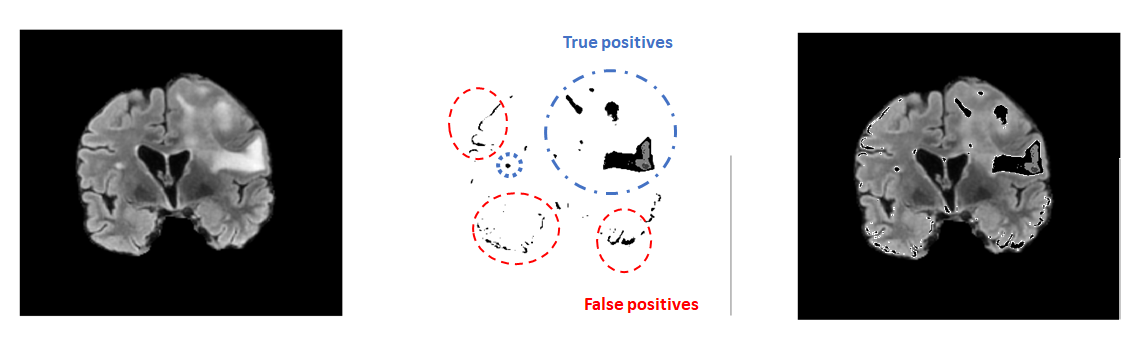}
\caption{A T1-weighted image of a brain cancer (left), the segmented image by the proposed method (middle), and the overlaid image (right) }\lel{Fig:RSNA3}
\end{figure}
}\fi
Figure \r{Fig:RSNA3} reports a T1-weighted MR image of different brain tumors. The most striking difference between the T1-weighted and FLAIR images is the white matter (WM) that is the exterior part inside the skull. In the T1-weighted MR image, the WM is very bright -- but still less bright than tumors -- and displays a numeric figure between 0 and 1 as its pixel value  while it is dark gray in the FLAIR image with its pixel value being almost 0. Therefore, the existence of the WM in the T1-weighted MR image tends to render successful segmentation of tumors only more challenging. This is why the FLAIR image is preferred for the segmentation. As already noticed, the T1-weighted image in Figure \r{Fig:RSNA3} shows more FPS pixels. A point worth noting here is that the proposed method successfully segmented the tumor only -- which is in the small blue circle in the middle figure -- even in the T1-weighted image; this closely accords with the fact that the WM is less bright than the tumor, thereby correctly specified as a normal cell by the proposed method. Even though the performance of the image segmentation deteriorated in the T1-weighted image, this issue can be easily resolved in that both the T1-weighted and FLAIR images from the same patient are simultaneously tried for the segmentation of tumors, and hence, only commonly segmented pixels are used for detecting tumors. During this process, the issue of the FPS pixels due to the WM can be alleviated to a great extent. Unfortunately, the dataset from Baid et al.\,\,\cite{Baid2021}, however, does not provide the FLAIR and T1-weighted images together, and hence, a further analysis is not viable.

There are other types of MR images such as T2-weighted and quantitative susceptibility mapping (QSM) images. Using the T2-weighted and QSM images together with those two other images will further enhance the performance of the proposed method.
\section{Conclusion}
This paper demonstrates the MD estimation methodology is versatile in that it can be applied to image segmentation problems, thereby extending its domain of application from traditional statistical problems to applied problems. This paper confines the investigation to the case of $K=2$ only, i.e., there exist two regions ($S_{1}^{\dagger}$ and $S_{2}^{\dagger}$) to segment. Investigation of the case of $K\geq 3$ will be an extension of findings in this study and form future research.

\bibliographystyle{plain}
\bibliography{MyRef}

\edt